\documentclass{ar-1col}
\usepackage{url}
\usepackage[numbers]{natbib}
\usepackage[utf8]{inputenc}
\usepackage[T1]{fontenc}
\usepackage{url}
\usepackage[margin=3cm]{geometry}
\usepackage{verbatim}

\begin{document}

\markboth{Barbeau et al.}{COHERENT at the SNS}

\title{COHERENT at the Spallation Neutron Source}

\author{P.S. Barbeau,$^{1,2}$ Yu. Efremenko,$^{3,4}$ and K. Scholberg$^1$
\affil{$^1$Department of Physics, Duke University, Durham, USA, 27708}
\affil{$^2$Triangle Universities National Laboratory, Durham, USA, 27708}
\affil{$^3$Department of Physics and Astronomy, University of Tennessee, Knoxville, USA, 37996}
\affil{$^4$Oak Ridge National Laboratory, Oak Ridge, USA, 37831}}

\begin{abstract}
The Spallation Neutron Source (SNS) at Oak Ridge National Laboratory provides an intense,  high-quality source of neutrinos from pion decay at rest.  This source was recently used for the first measurements of coherent elastic neutrino-nucleus scattering (CEvNS) by the COHERENT collaboration, resulting in new constraints of beyond-the-standard-model physics.   The SNS neutrino source will enable further CEvNS measurements, exploration of inelastic  neutrino-nucleus interactions of particular relevance for understanding of supernova neutrinos, and searches for accelerator-produced sub-GeV dark matter.   Taking advantage of this unique facility, COHERENT's suite of detectors in ``Neutrino Alley'' at the SNS is accumulating more data to address a broad physics program at the intersection of particle physics, nuclear physics, and astrophysics.  This review describes COHERENT's first two CEvNS measurements, their interpretation, and the potential of a future physics program at the SNS.

\end{abstract}

\maketitle

\tableofcontents






\section{INTRODUCTION}

The decays of positive pions stopped in a material produce a well-understood spectrum of multiple flavors of neutrinos ($\nu_\mu$, $\bar{\nu}_\mu$ and $\nu_e$) with energies up to half the muon rest mass. 
Such neutrinos may be used for studies of neutrino properties, neutrino-nucleus interactions, and probes of nuclear properties.  Recently, pion decay-at-rest neutrinos produced at the Spallation Neutron Source (SNS) at Oak Ridge National Laboratory (ORNL) in Tennessee have been used by the COHERENT collaboration for the first measurements of coherent elastic neutrino-nucleus scattering (CEvNS)~\cite{Akimov:2017ade,COHERENT:2020iec}, a process sought for decades using lower-energy  neutrinos produced at nuclear reactors.  
The SNS source of neutrinos also has a pulsed time structure, which enables robust background suppression. COHERENT's CEvNS measurements are the first of many; a suite of additional measurements  using low-energy recoil detectors with different nuclei will map out the nuclear target dependence of the CEvNS process.  The same neutrino source will enable measurements of inelastic interactions of neutrinos with nuclei.  Furthermore, COHERENT's detectors are  sensitive to distinctive signals from dark matter or other new particles produced in the SNS target.  The aim of this review is to describe the first CEvNS measurements as well as the suite of particle and nuclear physics experiments possible at this high-quality and readily accessible facility.

\section{NEUTRINO FLUXES FROM STOPPED PIONS}

\begin{figure}[h]
\includegraphics[width=3.6in]{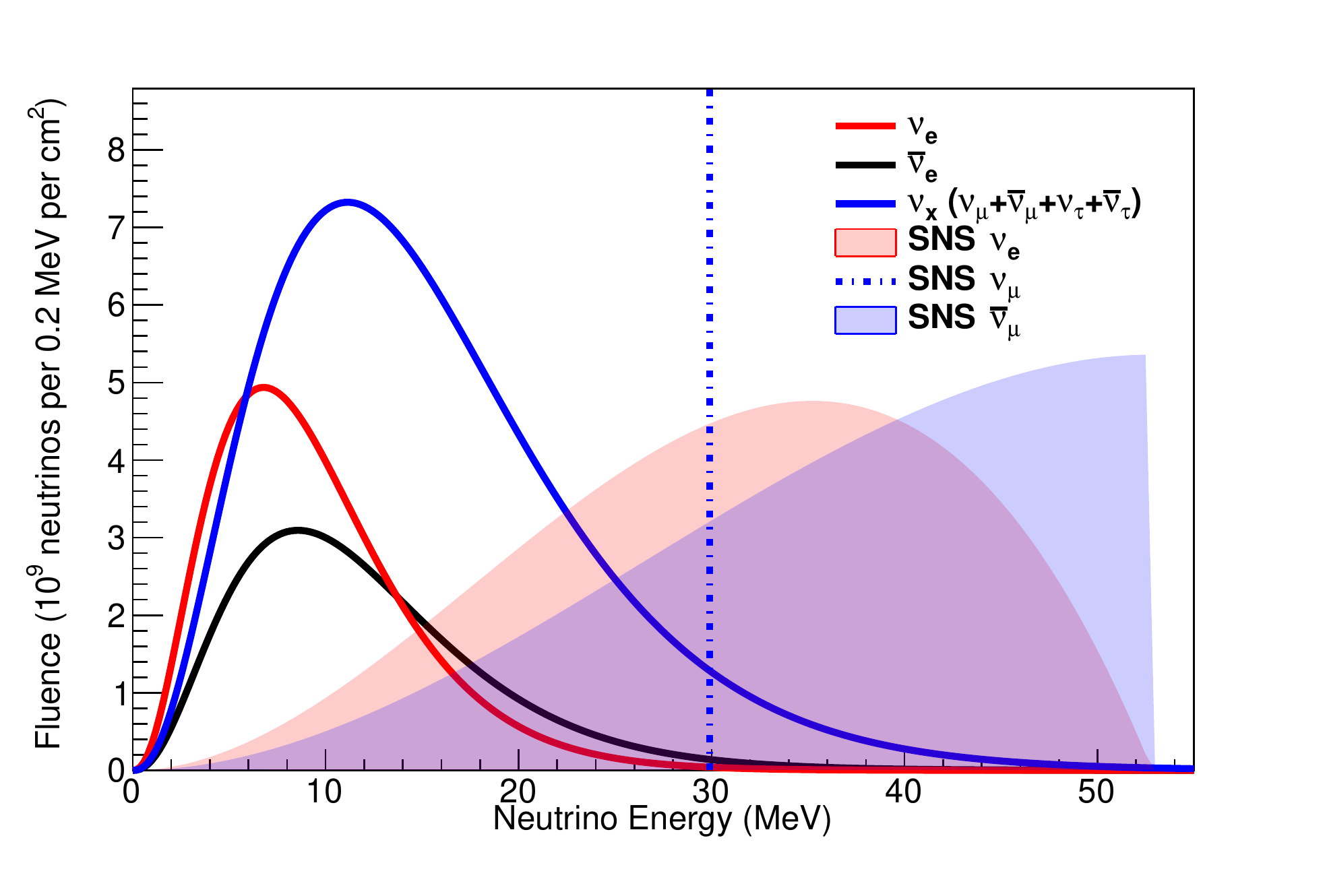}
\caption{Fluence spectrum of decay-at-rest neutrinos of each flavor at the SNS  at 27.5~m (COHERENT's argon detector distance) integrated over one day, compared with a typical supernova neutrino spectrum, integrated over the burst, for a core collapse at 10~kpc from Earth.  The supernova spectral parameters follow the nominal model in Ref.~\cite{GalloRosso:2017mdz}.}  
\label{fig:sn_sns_fluxes}
\end{figure}

The production of neutrinos from pions begins with the acceleration of protons to high energies (hundreds of MeV to GeV scale), that are then directed to collide with a target in order to produce copious secondary hadrons.  Protons with energies greater than $\sim$300 MeV will produce large numbers of pions, and the decays of these pions then produce neutrinos.  If high-energy (greater than a few hundred MeV) neutrinos are desired, 
pions from GeV-scale protons can be focused
and channeled into a decay pipe, so that they will decay in flight and transfer some of the pion momentum to the emitted neutrino; this method has been widely used in neutrino experiments, including long-baseline oscillation experiments~\cite{deGouvea:2013onf}.  However if the pions produced by the proton collisions lose energy in dense material, then they may stop and decay after coming to rest.   In dense material, negative pions are then captured by nuclei with a high probability.  As a result, the dominant neutrino production from stopped pions is from the weak-interaction decay, $\pi^+\rightarrow \nu_{\mu} + \mu^+$, followed by the decay at rest of the muon, $\mu^+ \rightarrow e^+ + {\nu_e} + \bar{\nu}_\mu$. 
The pion decay is fast, $\tau=26.033$~ns, whereas the muon decay time is 2.197~$\mu$s.    The shape of the spectrum of neutrinos from stopped pions is well known and is shown in Fig.~\ref{fig:sn_sns_fluxes}.   The prompt $\nu_\mu$ is monochromatic with energy 29.792~MeV, and the other flavor spectra range up to $m_\mu/2=52.8$~MeV.  Neglecting small radiative corrections, the $\nu_e$ spectral shape is described by $\phi(E_\nu) = 12 a (a E_\nu)^2(1-a E_\nu)$, and the $\bar{\nu}_{\mu}$ spectral shape is $\phi(E_\nu) = 2 a (a E_\nu)^2(3-2 a E_\nu)$, where $a=2/m_\mu$.

The optimal stopped-pion neutrino production occurs for proton energies between 650 and 1500~MeV~\cite{Alonso:2010fs}.
A clean stopped-pion spectrum that is free of contaminants is achieved for proton energies of about 1~GeV or less; such lower-energy protons will suppress a decay-in-flight component. 
In addition, a dense target for the protons is beneficial in order to maximize the likelihood that the pions stop and then decay at rest. 
Kaons can be produced if proton energies above $1.1$~GeV are used, resulting in high-energy neutrino spectral components. 

Stopped-pion neutrinos have been used in the past for several neutrino experiments, and others are currently in use and have been proposed~\cite{Athanassopoulos:1996ds,  Maschuw:1998jf,Brice:2013fwa,Baxter:2019mcx,CCM:2021leg,Alonso:2010fs}.
Here, we focus on the SNS, which has numerous favorable properties, among them a high neutrino flux and tightly pulsed timing.

\section{THE SPALLATION NEUTRON SOURCE} 

The Spallation Neutron Source (SNS) at Oak Ridge National Laboratory, constructed by
the United States Department of Energy's Office of Basic Energy Sciences, started its operation in 2007 with the goal of producing neutrons for multiple, diverse science topics, from materials science to fundamental neutron physics~\cite{Mason:2000wb}. 
The high-current proton accelerator started at 0.94 GeV proton energy and 0.8 MW beam power. Presently, the SNS is consistently running at 1 GeV proton energy and 1.4 MW beam power. By 2024 after the next round of upgrades, it will be running with 1.3 GeV proton energy and 2 MW beam power. 

The SNS First Target Station (FTS) proton beam~\cite{Henderson:2014paa} consists of a linear H$^-$ ion accelerator, an accumulator ring, and a proton target~\cite{Haines:2014kna}. The proton target employs liquid mercury contained inside a double-walled stainless steel vessel. Hydrogen ions accelerated in the 254-meter-long linear accelerator are stripped from their two electrons by a thin foil before entering the accumulator ring. This allows protons to change their direction of rotation in the magnetic field and smoothly enter the orbit of the accumulator ring. About 1000 initial proton pulses are accumulated in the ring to create short bunches of $\sim 10^{14}$  protons each, which are directed to the mercury target.  
The time required for protons to make a full loop in the 262-meter-long accumulator ring sets the maximum duration of the proton pulse on target. The SNS generates 400-nanosecond bursts of protons on target at 60 Hz frequency which allows for a highly effective suppression of backgrounds (factor of a few times $10^3$) and allows the simultaneous measurement of neutrino signal and backgrounds. The inner wall of the mercury containment vessel is continually eroded by cavitation, resulting in a need to replace it about twice per year.

Neutrons generated in the target are cooled down by room-temperature water and by cryogenic (20~K) hydrogen moderators and delivered to neutron beam lines on the target hall floor. An excellent location, dubbed ``Neutrino Alley,'' for neutrino experiments exists in the SNS basement that is well protected from the beam lines and the copious quantity of neutrons produced by the facility. It has about 70 meters of water-equivalent line-of-sight shielding to the SNS target and about 10 meters of water-equivalent overburden. The latter shielding is sufficient to eliminate the hadronic component of cosmic rays.

Detailed simulations of neutrino flux production in the SNS target are described in Ref.~\cite{Akimov:2021geg}, along with a validation of the overall flux normalization uncertainty of 10\% using existing hadron-production data sets.   The expected neutrinos per flavor produced per proton on target varies between about 0.15 and 0.4 for incident protons between 0.775 and 1.425~GeV.  The total predicted flux for 1~GeV protons at the FTS running at 1.4~MW is $4.7 \times 10^7$ cm$^{-2}$ s$^{-1}$ at 20~m from the target, with $\sim$99\% of the total flux expected to originate from $\pi^+$ decay at rest.

\begin{figure}[h]
  \begin{minipage}[c]{0.5\textwidth}
\includegraphics[width=2.5in]{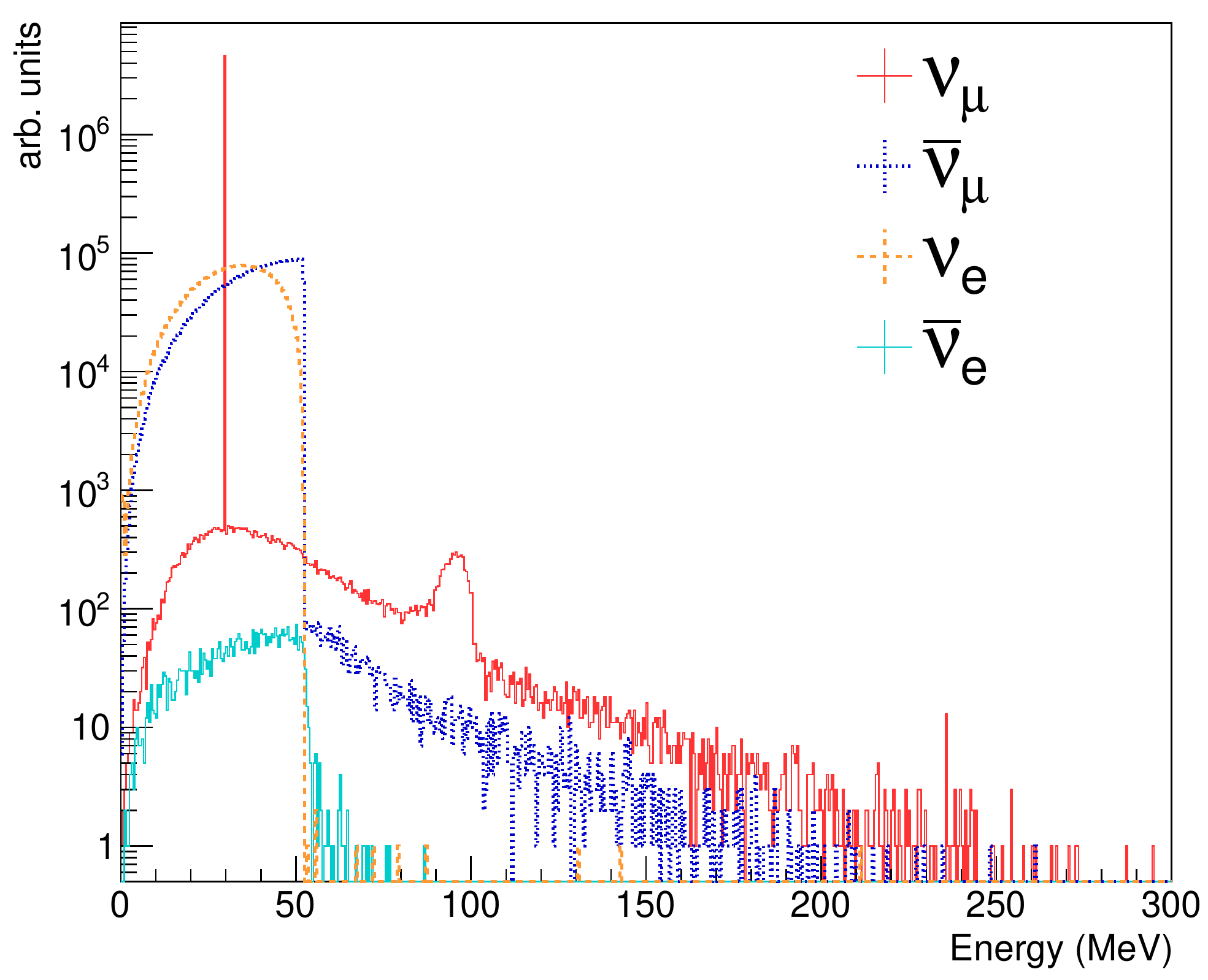}
  \end{minipage}\hfill
  \begin{minipage}[c]{0.5\textwidth}  
\includegraphics[width=2.5in]{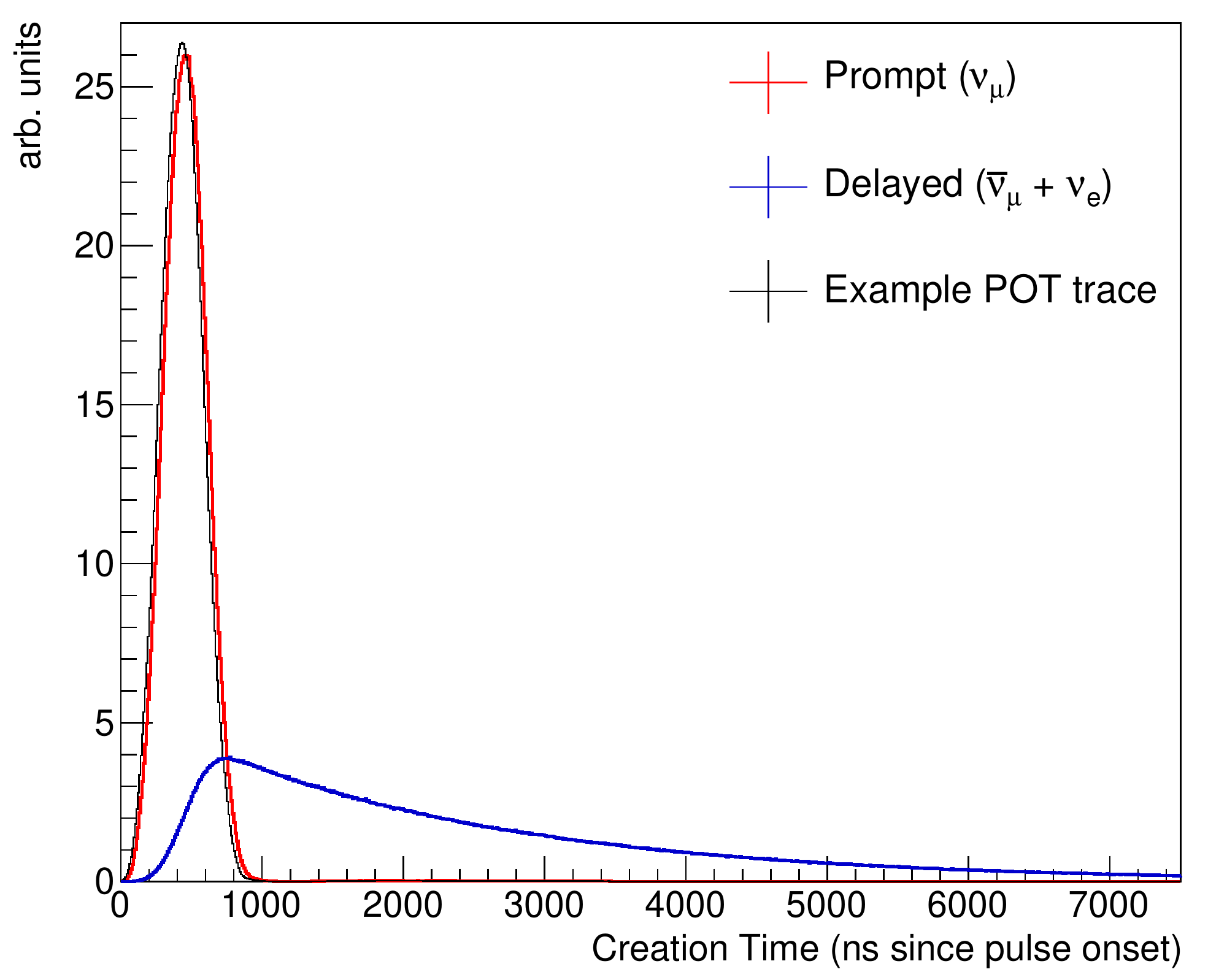}
  \end{minipage}\hfill
\caption{Left: energy distributions of the different flavor components from the COHERENT neutrino flux simulation.  Right: time distribution of prompt (pion decay) and delayed (muon decay) components. The thin black line shows the time structure of protons on target.  Figures from Ref.~\cite{Akimov:2021geg}.}
\label{fig:fluxsim}
\end{figure}

Interest in the use of the SNS as a neutrino source has a long history.  Prior proposals include ORLaND~\cite{Avignone:2000vh}, NuSNS~\cite{Stancu:2006hs},  OscSNS~\cite{Elnimr:2013wfa}, and CLEAR~\cite{Scholberg:2009ha}.

A second target station (STS) with a solid tungsten target is planned for the SNS~\cite{stsnote}.  For this stage, the total beam power will be increased to 2.8 MW and the proton beam will be split between two targets with 45 Hz to the first target and 15 Hz to the second, creating even more favorable conditions to suppress steady-state backgrounds. The estimated STS neutrino production per proton on target is slightly larger than for the FTS~\cite{Akimov:2021geg}.
A new hall for neutron instrumentation will be built for the STS, and sites for neutrino experiments in between the FTS and the STS, which will be irradiated by the isotropic neutrinos from both sources, are being explored.

\section{COHERENT ELASTIC NEUTRINO-NUCLEUS SCATTERING}\label{sec:cevns}

The coherent elastic scattering of neutrinos off nuclei (CEvNS) occurs when a neutrino scatters off an entire nucleus,
transferring some of its momentum to the nucleus as a whole, but creating no internal excitations of the nucleus or ejected particles. The ``elastic'' qualifier in this context means that no new particles are created and the final nucleus is in the ground state.\footnote{The process is not to be confused with e.g., coherent pion production~\cite{Higuera:2014azj}, which occurs at $\sim$GeV neutrino energies and is accompanied by creation of a pion in the final state.}
 The process is a weak neutral-current one, occurring via exchange of a $Z^0$ boson.  
We can think of the process as a scattering in which target nucleon wavefunctions
remain in phase with each other before and after the collision. The approximate condition for coherence is that $QR<<1$, where $Q$ is the momentum transfer to the nucleus, $R$ is the nuclear radius, and $M$ is the mass of the nucleus.  For medium-sized nuclei, this condition holds reasonably well for scattering of neutrinos with energies less than a few tens of MeV, such that a dominant fraction of interactions will preserve coherence.  

In the SM, the differential CEvNS cross section can be written as~\cite{Freedman:1977xn}:

\begin{equation}
\frac{d\sigma}{dT} = \frac{G_F^2 M}{2\pi} F^2(Q)\left[(G_V+G_A)^2 + (G_V-G_A)^2\left(1-\frac{T}{E_{\nu}}\right)^2 - (G_V^2-G_A^2)\frac{MT}{E_{\nu}^2}\right]. 
\end{equation}

Here,   $T$ is the recoil energy of the nucleus, $E_{\nu}$ is the incident neutrino energy, and $G_F$ is the Fermi constant.
 $F(Q)$ is the nuclear form factor as a function of momentum transfer $Q=\sqrt{2MT+T^2}\sim \sqrt{2MT}$.  The vector coupling constant is $G_V=g_V^pZ + g_V^nN$,
where
$Z$ is the number of protons in the target nucleus, $N$ is the number of neutrons, $g_V^p \sim (\frac{1}{2}-2\sin^2 \theta_W)$, $g_V^n \sim -\frac{1}{2}$, and $\theta_W$ is the effective weak mixing angle (omitting well-understood radiative corrections here).
The axial coupling constant $G_A$ is given by a similar expression, with $Z$ and $N$ replaced by net numbers of unpaired nucleons, and with $g_A^p$ and $g_A^n$ having opposite sign for neutrinos and antineutrinos. 
However, because the total number of unpaired nucleons is typically much smaller than  the total number of protons and neutrons in a nucleus, axial contributions are neglected  (although they are potentially of interest for future precision CEvNS experiments--- see Sec.~\ref{sec:nai}).
Taking into account radiative corrections and the neutrino charge radius~\cite{Erler:2013xha}, $g_V^p =  0.0298$ and $g_V^n = -0.5117$.

For $T<<E_\nu$, the differential cross section can be rewritten to a good approximation as
\begin{equation}
\left(\frac{d\sigma}{dT}\right)_{\nu A} = \frac{G_F^2}{2\pi} \frac{Q_W^2}{4} F^2(Q) M \left[2 - \frac{M T}{E_\nu^2}\right],
\end{equation}

where $Q_W$ is the weak charge of the nucleus,  $Q_W = N-(1-4\sin^2\theta_W) Z$. Because $4 \sin^2 \theta_W \sim 1$, the term proportional to $Z$ in $Q_W$ is small; therefore the weak charge of the nucleus is roughly proportional to $N$ and the CEvNS cross section is roughly proportional  to $N^2$ for a given nucleus.

The $Q$-dependent nuclear form factor is the Fourier transform of the nucleon spatial distribution within the nucleus.  It can be reasonably approximated by several different functional forms~\cite{Piekarewicz:2016ic}; the choice of form factor description results in percent-level or smaller differences for $Q$ values of relevance for stopped-pion neutrinos.  The form factor differs in general for protons and neutron constituents of the nucleus, and can also differ for vector and axial contributions to the CEvNS cross section, although to first approximation these form factors can be considered identical.
For a fully coherent interaction, $F=1$.  The smaller the $Q$ for a given interaction, the smaller the form-factor suppression.   COHERENT uses the Klein-Nystrand parameterized functional form~\cite{Klein:1999qj}, 
$|F(Q)|^2= \left(\frac{3(\sin(QR)-QR \cos(QR))}{(Q R)^3(1+a_{kn}^2 Q^2)}\right)^2$, where $R = 1.2 (N+Z)^{1/3}$, $a_{kn}=0.7$,
to describe the form factor, which gives negligibly different results from the popular Helm form factor~\cite{Helm:1956dv}. 

\begin{figure}[h]
\includegraphics[width=3in]{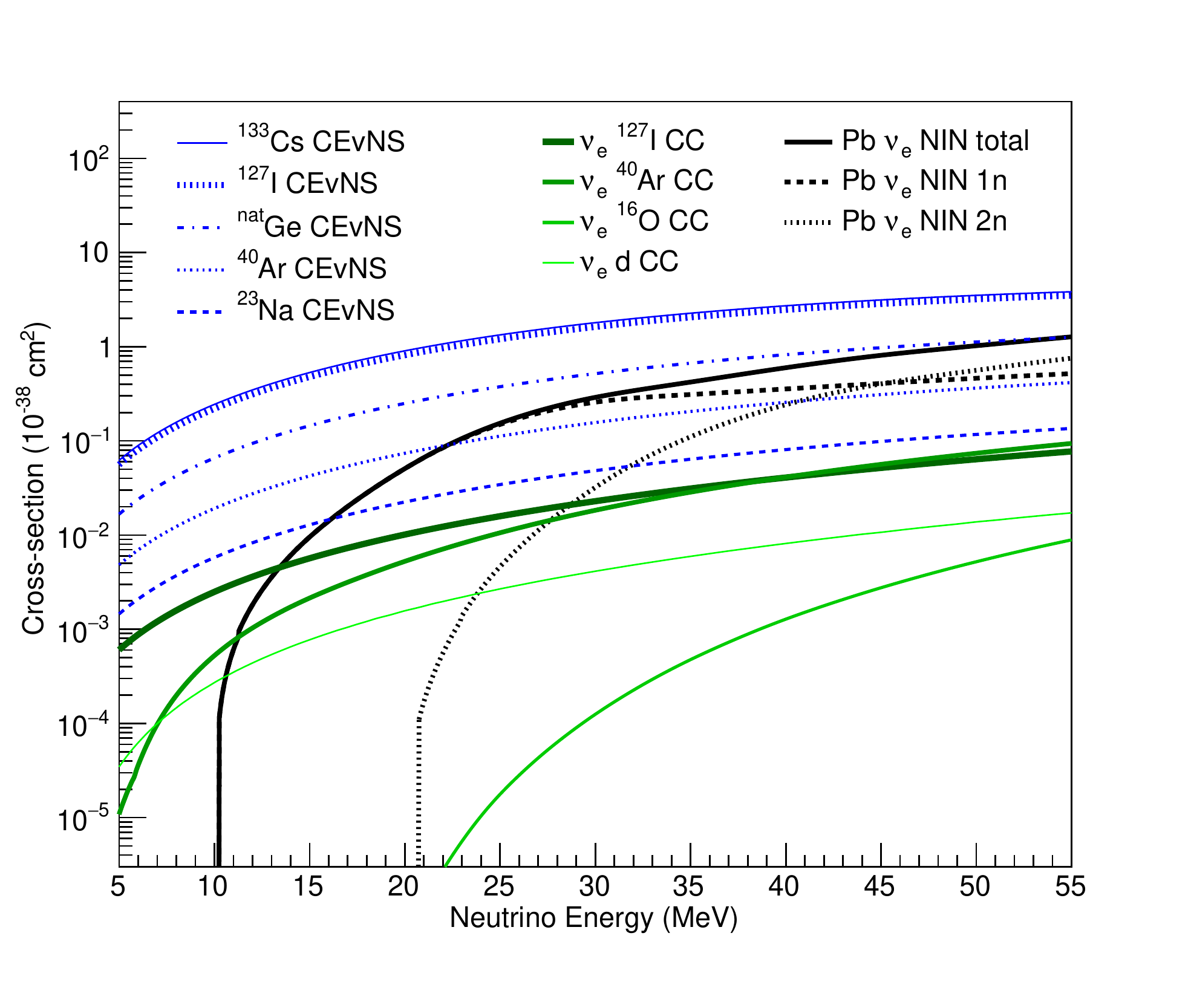}
\caption{Left: Total CEvNS cross section as a function of energy for several nuclear targets relevant for COHERENT are shown in blue.  Inelastic cross sections (see Sec.~\ref{sec:inelastics}, ~\cite{snowglobes}) are shown in green, and neutrino-induced-neutron (NIN) cross sections (see Sec.~\ref{sec:nins}) are shown in black.}
\label{fig:xscns}
\end{figure}

For neutrino energies less than about $\sim$100 MeV, although charged-current (CC) and neutral-current (NC) neutrino-nucleus interactions do occur (see Sec.~\ref{sec:inelastics}), the CEvNS interaction channel dominates the neutrino-nucleus cross section for most nuclei.  Fig.~\ref{fig:xscns} shows neutrino-interaction cross sections of relevance to COHERENT.

\subsection{Detection of CEvNS}

The existence of the CEvNS process was first posited by Freedman in 1973~\cite{PhysRevD.9.1389}, with a paper by Kopeliovich and Frankfurt~\cite{Kopeliovich:1974mv} following shortly thereafter.  Freedman declared the proposed detection of CEvNS an ``act of hubris,'' due to the associated ``grave experimental difficulties.''  Although, as for all neutrino interactions, the interaction cross section is tiny,  CEvNS interactions are not (relatively speaking) rare, due to the $\sim N^2$ enhancement of the cross section. Thus, the experimental difficulties do not arise due to a small cross section.  The issue is, rather, the kinematics of the interaction: the resulting kinetic energy $T$ of the recoiling nucleus is very small.  The maximum recoil energy to the nucleus is given by 
$T_{\rm max} = \frac{2 E_\nu^2}{M+2 E_\nu}$.
For medium-sized nuclei, this amounts to several tens of keV for incident neutrinos in the few tens of MeV range.  

Such low-energy signals are challenging to measure.  For typical neutrino detectors, thresholds for particle detection are at least several MeV.  However, in recent decades there has been a significant development of technologies designed specifically to detect tiny nuclear recoils, mostly in the context of weakly interacting massive particle (WIMP) dark matter searches, for which the signal is exactly the tiny recoil of a jostled nucleus.    The recoil energy can be collected in various ways: as photons, as ionized charge, as phonons, or a material phase change due to added heat.  Detectors can in principle use energy coupled to more than one detection channel, and use the relative amounts in different channels as a nuclear recoil discriminant, provided there is enough recoil energy.  This strategy is popular for dark matter detectors~\cite{Gaitskell:2004gd}.  Some detectors can select nuclear recoils by other means, such as by analyzing the arrival time of prompt and delayed scintillation light via pulse shape discrimination (e.g.,~\cite{Boulay:2006mb}.)
Understanding of detector response to recoil energy deposition is critical for interpretation of experimental data.  In many cases, recorded energy deposition for nuclear recoils is suppressed ("quenched") with respect to that for electron or gamma energy deposition, by a recoil-energy-dependent factor known as the ``quenching factor (QF)''--- see Sec.~\ref{sec:qf}.\footnote{For detector materials in which observable nuclear recoil energies depositions are quenched,  true recoil energies $T$ are denoted with energy units "r," e.g., keVr, whereas electron-equivalent energies are denoted by "ee," e.g., keVee.  The ratio of observed electron-equivalent energy to recoil energy is given by the quenching factor, i.e., $E_{ee}= ({\rm QF}) T$.} 

For detection of CEvNS, one needs an appropriate low-energy neutrino source.  Reactors have long been the sources of choice for CEvNS searches, thanks to their copious neutrino  fluxes.  However, given that typical reactor neutrino energies are several MeV, one expects sub-keV nuclear recoil energies, which are difficult to detect even for many sensitive technologies.  Furthermore, the total CEvNS cross section scales as $\sim E_\nu^2$.  Therefore, higher energies are beneficial, up until the point at which CEvNS is strongly suppressed relative to inelastic scattering (i.e., for which the neutrino is predominantly scattering off nucleons rather than entire nuclei.) 
Neutrinos from stopped pions balance several trade-offs and are almost optimal for studying CEvNS, in that the CEvNS rate is high and recoil energies are more easily detectable above threshold, yet a high fraction of coherence is maintained. 
The modest lost of coherence at stopped-pion energies can also be valuable for understanding of nuclear structure, given that one can probe the form factor for various nuclei as a function of $Q$.

\section{PHYSICS REACH OF CEVNS DETECTION AT THE SNS}

The CEvNS cross section is cleanly predicted in the standard model (SM) of particle physics--- nuclear structure effects are present via the form factor, but are known to the few percent level.  Given the relatively small theoretical uncertainty, a CEvNS measurement represents a test of SM weak physics.  Any deviation from the SM prediction in the rate or recoil spectrum could be an indication of new contributions to the interaction cross section.  We can divide potential new physics signatures into two generic categories: those which affect the \textit{total rate} and those which affect the \textit{shape of the recoil spectrum}.  The latter in particular are more vulnerable to systematic uncertainties on the energy dependence of the detection response.  

In general, the experimental observables that can distinguish SM from beyond-the-standard-model (BSM) signals are: the observed recoil time with respect to neutrino production, the absolute event rate on a given target nuclear species, the recoil energy distribution, and, in principle, the angular distribution of recoils.  Other parameters that can be modified by experimentalists and for which the dependence of event rates can be measured are: the baseline (distance traveled from the source) and the angle of the flight path with respect to the beam axis, as well as the $N$ and $Z$ of the target nuclear species.

In the following we will explore several physics topics accessible using measurements of CEvNS using a stopped-pion source. Table~\ref{tab:physics1} summarizes these physics opportunities for COHERENT.  In Section~\ref{sec:results}, we will present some specific interpretations using COHERENT data.

\begin{table*}

\begin{tabular}{c|c|c}
Topic & Experimental signature & Detector requirements \\ \hline \hline
Non-standard neutrino  & Deviation from $N^2$,&  Multiple targets,\\
interactions, new mediators & deviation from SM recoil shape, & energy resolution,\\ 
 & event rate scaling & quenching factor \\ \hline
Weak mixing angle &  Event rate scaling & Multiple targets,\\
   & & quenching factor \\ \hline
Neutrino magnetic moment & Low recoil energy excess& Low energy threshold,\\
 & & energy resolution, \\
 & & quenching factor \\ \hline
Inelastic CC/NC cross-section & High-energy (MeV) & Large mass\\
for supernova & electrons, $\gamma$s & \\ \hline
Inelastic CC/NC cross-section & High-energy (MeV) & Large mass\\
for weak coupling parameters & electrons, $\gamma$s & \\ \hline
Nuclear form factors & Recoil spectrum shape& Energy resolution,\\
  & & multiple targets, \\ 
  & & quenching factor \\ \hline
Accelerator-produced dark matter & Event rate scaling, & Energy resolution,\\
& recoil spectrum shape, & quenching factor \\
& timing, direction & \\
& with respect to source  & \\ \hline
Sterile oscillations & Event rate and spectrum & Similar or movable\\
   &  at multiple baselines & detectors at different\\ 
     &  & baselines \\ 
 
 \end{tabular}
\caption{COHERENT physics topics, and corresponding experimental requirements. ``Quenching factor'' refers to the requirement to understand detector response for nuclear recoils.}\label{tab:physics1}
\end{table*}

\subsection{Standard-Model Weak Mixing Angle}

In the context of the SM, the measured CEvNS event counts can be used to infer the weak charge, $Q_W = N - (1-4 \sin^2 \theta_W) Z$, of the nucleus, and hence $\theta_W$.  A measurement that differs from the expected dependence of the effective weak mixing angle on $Q$ in the low-energy regime could be a signature of extra neutral weak currents or supersymmetric extensions to the Standard Model~\cite{Safronova:2017xyt}.  We note an existing experimental anomaly with respect to the SM for neutrino-nucleon scattering at the $Q\sim$~GeV~c$^{-1}$ scale~\cite{Zeller:2001hh}. The current uncertainty from CEvNS measurements is at the $\sim$10\% level at stopped-pion $Q$ values of a few tens of MeV~c$^{-1}$; this is not currently competitive with other methods for determining $\theta_W$ at low $Q$ from parity-violating electron-proton scattering~\cite{Androic:2018kni}, Moller scattering~\cite{Benesch:2014bas} and atomic parity violation~\cite{Roberts:2014bka}.  However, none of these experiments probe BSM physics that is specific to interactions of neutrinos and quarks, which would manifest as an inferred value of $\theta_W$ differing from the nominal SM expectation.  Such possibilities of new interactions of neutrinos and quarks are considered in the following Sec.~\ref{sec:nsi}.

\subsection{Non-standard Interactions of Neutrinos (NSI)}\label{sec:nsi}

As one example of a test of BSM physics, we consider a new vector coupling mediated by heavy particles which results in an overall scaling of the CEvNS cross section. A standard  parameterization of such a new flavor-dependent interaction uses $\varepsilon$ couplings~\cite{Barranco:2005yy,Scholberg:2005qs}. We assume that that spin-dependent axial NSI contributions are small.  The dominant
vector couplings are denoted $\varepsilon_{\alpha \beta}^{qV}= \varepsilon_{\alpha
\beta}^{qL}+\varepsilon_{\alpha \beta}^{qR}$, for quark flavor $q$, initial-state neutrino flavor $\alpha$ and final-state neutrino flavor $\beta$.  
In this formulation, the differential CEvNS cross-section with NSI is described as:

\begin{eqnarray}\label{eq:xscn}
\lefteqn{\left(\frac{d\sigma}{dT}\right)_{\nu_\alpha A} = 
  \frac{G_F^2 M}{\pi} F^2(2MT)\left[1 - \frac{M T}{2E_\nu^2}\right] \times} \\
&     \{[Z(g_V^p + 2\varepsilon_{\alpha \alpha}^{uV}+ \varepsilon_{\alpha \alpha}^{dV})+ N(g_V^{n} + \varepsilon_{\alpha \alpha}^{uV}+ 2\varepsilon_{\alpha \alpha}^{dV})]^2 & \nonumber \\ 
 &            + \displaystyle \sum_{\alpha \ne \beta }{[Z(2\varepsilon_{\alpha \beta}^{uV}+ \varepsilon_{\alpha \beta}^{dV})+ N(\varepsilon_{\alpha \beta}^{uV} + 2 \varepsilon_{\alpha \beta}^{dV})]^2 \}}. & \nonumber 
 \end{eqnarray}
 The effect of non-zero values of $\varepsilon$s, which can be either positive or negative, can be either an enhancement or suppression of the CEvNS rate; some combinations of NSI parameter values for a given $Z$ and $N$ can result in the SM CEvNS rate.  A combination of CEvNS measurements on targets with different $N$ and $Z$ values can break any such accidental degeneracies (as well as cancel flux-related uncertainties.)

NSI parameters can therefore be probed by measuring the rate of CEvNS and comparing to SM expectation.  A stopped-pion source has both electron and muon flavor content; hence a CEvNS measurement gives direct access to all but $\varepsilon_{\tau \tau}$ couplings.  Furthermore, because at the SNS neutrino flavors can be separated by timing, one can probe electron and muon NSI separately.

Degeneracies in neutrino oscillation parameter measurements can occur if NSI are allowed to exist~\cite{Coloma:2016gei, Coloma:2017egw}, and CEvNS measurements can help to resolve these degeneracies.
For example, the ``LMA-Dark'' degeneracy can confound determination of the mass ordering at long-baseline experiments~\cite{Coloma:2017ncl}, and CEvNS measurements can help resolve the ambiguity.

The COHERENT detector setup, which will be described in the following sections, will enable searches for non-zero NSI.  An observed result consistent with the SM will result in new constraints on these and other $\varepsilon$ parameters. 

NSI with BSM light mediators, of mass comparable to $\sqrt{Q}$, will result in a distortion of the recoil spectrum $T$~\cite{Liao:2017uzy} with respect to SM expectation.
Following the discussion in Ref.~\cite{Liao:2017uzy}, one substitutes
$Q_w^2=Z g_p^V+ N g_n^V$  with a non-standard $Q_{\alpha, NS}^2$ according to 

\begin{equation}
Q_{\alpha,\rm{NSI} }^2=\left[Z\left(g_p^V+\frac{3g^2}{2\sqrt{2}G_F(Q^2+M_{Z'}^2)}\right)
+N\left(g_n^V+\frac{3g^2}{2\sqrt{2}G_F(Q^2+M_{Z'}^2)}\right)\right]^2\,.
\label{eq:lightmed}
\end{equation}
Here $g$ is the new coupling and $M_{Z'}$ is the new mediator mass.  For sensitivity to such models, recoil energy resolution is desirable.

Further examples of the potential of NSI constraints from CEvNS can be found in Refs.~\cite{Coloma:2017ncl,Liao:2017uzy, Dent:2017mpr}. 

\subsection{Neutrino Electromagnetic Properties}

Although neutrino electromagnetic interactions are tiny in the SM, BSM-induced electromagnetic effects can be larger.  CEvNS provides a probe of these properties.

\subsubsection{Neutrino Charge Radius}
The neutrino charge radius has a small flavor-dependent effect on the CEvNS cross section.  Using a stopped-pion flux, percent-level measurements of muon- and electron neutrino CEvNS rates, which can be separated by timing at the SNS, enable a measurement of the effective neutrino charge radius~\cite{Papavassiliou:2005cs,Cadeddu:2018dux}.

\subsubsection{Neutrino Magnetic Moment}

Neutrino magnetic scattering results in a characteristic turn-up of the spectrum at low recoil energy, both for scattering on electrons and on nuclei~\cite{Vogel:1989iv,Dodd:1991ni,Scholberg:2005qs,Kosmas:2015sqa}.
For a spin-zero nucleus, the differential cross section as a function of recoil energy can be written~\cite{Vogel:1989iv}
\begin{equation}
\left(\frac{d\sigma}{dT}\right)_{\rm m} = \frac{\pi \alpha^2 \mu_\nu^2 Z^2}{m_e^2}\left(\frac{1-T/E_\nu}{T}+ \frac{T}{4E_\nu^2} \right),
\end{equation}
where $m_e$ is the mass of the electron, $\alpha$ is the fine structure constant, and $\mu_\nu$ is the neutrino magnetic moment in units of Bohr magnetons.
The SM predicts a very tiny magnetic moment of $\mu_\nu =
3.2\times 10^{-19} \mu_B (\frac{m_{\nu}}{1 {\rm eV}})$~\cite{Zyla:2020zbs}. Any measurement of magnetic moment larger than this would be a signature of BSM physics;  furthermore anomalously large magnetic moments would strongly hint that neutrinos are Majorana fermions~\cite{Bell:2006wi}.  The current best limits on neutrino magnetic moment have been set using neutrino-electron scattering of solar neutrinos by Borexino~\cite{Borexino:2017fbd}.   Indirect astrophysical limits are more stringent~\cite{Zyla:2020zbs}.  We note that a recent XENON experiment measurement has a potential interpretation as a non-zero neutrino magnetic moment from scattering on electrons~\cite{Aprile:2020tmw}, although other explanations exist for the observed effect.
A flux of stopped-pion neutrinos allows measurements with near-pure $\nu_\mu$ flavor~\cite{Scholberg:2005qs}, for which the most stringent magnetic moment limits are from LSND~\cite{Auerbach:2001wg, Kosmas:2015sqa}.  
For non-zero neutrino magnetic moment searches, low energy threshold and good energy resolution are desirable.

\subsection{Sterile Neutrino Oscillations}

Several interesting experimental hints exist that are compatible with the existence of sterile neutrinos, i.e., new neutrino states that have no SM weak interactions. Although there is no specific model fully compatible with all experimental data, the hypothesis remains viable, and a worldwide program is underway to explore this possibility via a range of experimental approaches~\cite{Abazajian:2012ys,Giunti:2019aiy}. 
CEvNS is an excellent tool for the search for sterile
neutrino oscillations, given that flavor-blind neutral-currents can probe the disappearance of active neutrinos unambiguously. Both a distortion of recoil spectra and an overall rate suppression can be used to test for the presence of sterile oscillations.  A setup involving  multiple detectors at different baselines, so that correlated uncertainties in the source can be canceled, is desirable.    Alternatively, a single detector could view multiple sources at different baselines~\cite{anderson:2012pn}, thereby canceling detector-related uncertainties.  The latter scenario could be achieved using a single detector strategically sited between the FTS and STS. Sensitivity is statistics-limited, such that typically tens-of-tonne-scale detectors are required to fully cover parameter space of interest~\cite{anderson:2012pn,Kosmas:2017zbh, Blanco:2019vyp}.

\subsection{Nuclear Form Factors}

As described in Sec.~\ref{sec:cevns}, the form factor $F(Q)$ describes the spatial distribution of nucleons in the nucleus.
The observed recoil energy $T$ determines $Q$; therefore, the observed CEvNS recoil energy spectrum allows one to map the effect of a non-unity form factor (i.e., finite spatial size) of the nucleus.    Since the nuclear weak charge is strongly dominated by its neutron content, CEvNS is sensitive primarily to the neutron form factor of the nucleus.
Furthermore, since proton spatial distributions in nuclei are generally well understood, a measure of the mean radius of the neutron distribution (the ``neutron radius'') enables determination of the ``neutron skin'' of a nucleus --- the difference between the larger neutron radius and the proton radius.  These nuclear parameters are of interest for understanding of neutron stars~\cite{Reed:2021nqk}, and the most precise measurements of them so far are via parity-violating electron scattering~\cite{PREX:2021umo}.

For many nuclei, neutron distributions are already understood theoretically to the few percent level (e.g., for argon~\cite{Payne:2019wvy}); this is an advantage from the point of view that nuclear uncertainties are small and hence CEvNS is a clean probe of BSM physics.  However, if one assumes SM physics, then at high recoil-energy precision 
we will be able to use CEvNS measurements to \textit{measure} nuclear form factors~\cite{Amanik:2009zz,Patton:2012jr,Cadeddu:2017etk}, and infer the neutron radius and neutron skin depth.  With sufficiently precise measurements, axial-vector contributions to the weak nuclear response can be determined, in principle, for nuclei with spin~\cite{Hoferichter:2020osn}.  Currently, the theoretical uncertainty on nuclear form factors is smaller than experimental precision.  However as CEvNS measurements improve, they will provide determination of nuclear neutron distributions independent from other methods.  We note that at sufficient experimental precision, a framework for disentangling nuclear response effects from possible BSM signatures will be required~\cite{Hoferichter:2020osn}.

\subsection{Accelerator-Produced Sub-GeV Dark Matter}
Another physics topic that can be addressed by an experiment at a stopped-pion source is a search for accelerator-produced dark matter in models
with ``portal'' particles that mediate interactions between relic DM and SM particles.  In such models, neutral mesons produced in the SNS target 
(pions and etas) produce portal particles via their interaction or decay, and these portal particles then decay to dark matter that can produce nuclear recoils in COHERENT detectors with specific recoil energy signatures~\cite{deNiverville:2015mwa,deNiverville:2016rqh,Dutta:2019nbn}.  
CEvNS events from neutrinos are the background for this search.

The time structure of the beam is especially valuable for separating SM CEvNS from DM-induced recoils.  Because the DM signal is expected to be prompt, the delayed CEvNS signal can provide a powerful constraint on the CEvNS background in the prompt window.  The DM signal will also have a distinctive dependence on direction with respect to the source, whereas CEvNS signals will be isotropic.
COHERENT's sensitivity to this signal for two benchmark models:  a vector portal particle with kinetic mixing to a  photon~\cite{Boehm:2003hm,Fayet:2004bw,deNiverville:2011it}, and a model with a leptophobic portal coupling to SM baryons~\cite{Batell:2014yra} is explored in Ref.~\cite{Akimov:2019xdj}.  Dark matter particle masses in the few to few hundred MeV c$^{-2}$ range can be probed.

\subsection{CEVNS at Dark Matter Direct Detection Experiments} 

There are deep connections between
CEvNS and WIMP (weakly interacting massive particle) detection experiments~\cite{Drukier:1983gj,Cabrera:1984rr}.
Searches for relic WIMP dark matter have been underway for the past several decades, making use of low-threshold, low-background detectors with a range of technologies.  The primary signature of WIMP dark matter in these detectors is a single nuclear recoil.  CEvNS from natural neutrino sources--- the Sun, diffuse supernova neutrino background, and atmospheric neutrinos, will produce a nuclear recoil signature which is identical to a WIMP scatter on an event-by-event basis. The SM CEvNS-induced background will have a predictable recoil spectrum, which can be distinguished from a WIMP-induced one, given sufficient statistics.  However, for WIMP parameters such that the event rate is low enough, the neutrino and WIMP recoil spectra are statistically indistinguishable, creating  the so-called ``neutrino floor'' for WIMP dark matter detection~\cite{Monroe:2007xp,Gutlein:2010tq,cushman2013:snowmassDM,anderson:2011cevnsDM, billard:2014}.  The CEvNS background will have also a distinct directionality, which motivates next-generation directional recoil detectors with the purpose of reaching below the neutrino floor~\cite{Vahsen:2020pzb}. 

Naturally occurring neutrinos also constitute a CEvNS \textit{signal} for WIMP dark matter detectors.  Solar neutrinos will be visible as an all-flavor signal in large detectors~\cite{Drukier:1983gj,Billard:2014yka}, as will neutrinos from supernova bursts~\cite{Horowitz:2003cz}.   For CEvNS as either astrophysical signal or dark matter background, a BSM-induced modification of the cross section would create difficulties in signal interpretation.  This motivates an independent laboratory-based CEvNS measurement with a stopped-pion source, which covers the neutrino energy range of interest for neutrino floor in the $10-10^3$ GeV $c^{-2}$ WIMP mass range.

The SNS stopped-pion source can also serve as a ``test beam'' for development of low-energy nuclear-recoil sensitive detector technology, given that the CEvNS events will have known recoil spectrum, direction, and timing.

\section{INELASTIC NEUTRINO-NUCLEUS INTERACTIONS}\label{sec:inelastics}

Inelastic charged-current (CC) and neutral-current (NC)	interactions are distinct from CEvNS.  CC neutrino-nucleus interactions can be either quasi-elastic, i.e., resulting only in the ground state of the final state nucleus, or they can also result in an excited final-state nucleus which then de-excites by gamma emission or the ejection of nucleons. Such nuclear debris is in principle observable in conjunction with the final-state lepton.   NC neutrino-nucleus interactions are neutrino-induced excitations for which de-excitation products are similarly observable. Only $\nu_e$ CC interactions are accessible with the SNS stopped-pion flux, given that $\nu_\mu$ and $\bar{\nu}_\mu$ are below CC threshold of $\sim$110~MeV, and the $\bar{\nu}_e$ flux from the SNS is very small.   Cross sections for inelastic neutrino-nucleus interactions are typically a couple of orders of magnitude smaller per target nucleus than CEvNS cross sections in the stopped-pion energy range; however, the interaction product energies are typically of the same order as the incident neutrino energy, and so around three orders of magnitude larger than CEvNS recoils.  The experimental requirements for CEvNS and inelastics are therefore quite different.  Larger masses are needed for inelastics, as well as the dynamic range to record MeV-scale energy depositions, while very low thresholds are not required. 

Inelastic neutrino-nucleus cross sections are so far quite poorly measured in the few-tens-of-MeV regime, with few measurements existing, none of which have better than around 10\% uncertainty~\cite{Formaggio:2013kya}.  Cross section measurements are of intrinsic interest for the study of the structure of neutrino-nucleus weak scattering (see Sec.~\ref{sec:nai}).   Furthermore, stopped-pion energies belong to the regime of direct relevance for supernova neutrino detection, as well as for understanding of processes inside a supernova involving neutrinos. The burst of neutrinos from a core-collapse supernova includes neutrinos of all flavors, with average energies of about 10-15~MeV and with tails ranging up to several tens of MeV~\cite{Scholberg:2012id}; see Fig.~\ref{fig:sn_sns_fluxes}. 

With the exception of cross sections on simple targets, such as neutrino-electron elastic scattering and inverse beta decay of $\bar{\nu}_e$ on free protons, theoretical understanding of low-energy neutrino interactions is also relatively poor, with uncertainties on the cross sections of up to a factor of a few.  The well-understood stopped-pion neutrino spectrum provided by the SNS is a near-ideal neutrino source for improved measurements.  Several of the COHERENT detectors will have sufficient mass to enable statistically meaningful measurements.
The ${}^{40}$Ar$(\nu_e, e^-){}^{40}$K cross section is of direct and primary relevance for supernova neutrino detection in the Deep Underground Neutrino Experiment~\cite{DUNE:2020zfm,Gardiner:2021qfr}.  The NaI[Tl] detector array can be used to observe CC reactions in $^{127}$I and $^{23}$Na. Finally, the heavy-water detector will be sensitive to CC ${}^{16}$O$(\nu_e,e^-){}^{16}$F on oxygen, using both light-water and heavy-water components.  These interactions are relevant for Super-K and Hyper-K (see Sec.~\ref{sec:d2o_det}). See Sec.~\ref{sec:nins} for discussion of neutrino-induced neutrons.

\section{THE COHERENT EXPERIMENTAL PROGRAM}

\begin{figure}[htbp]
\centering
\includegraphics[height=3.5in]{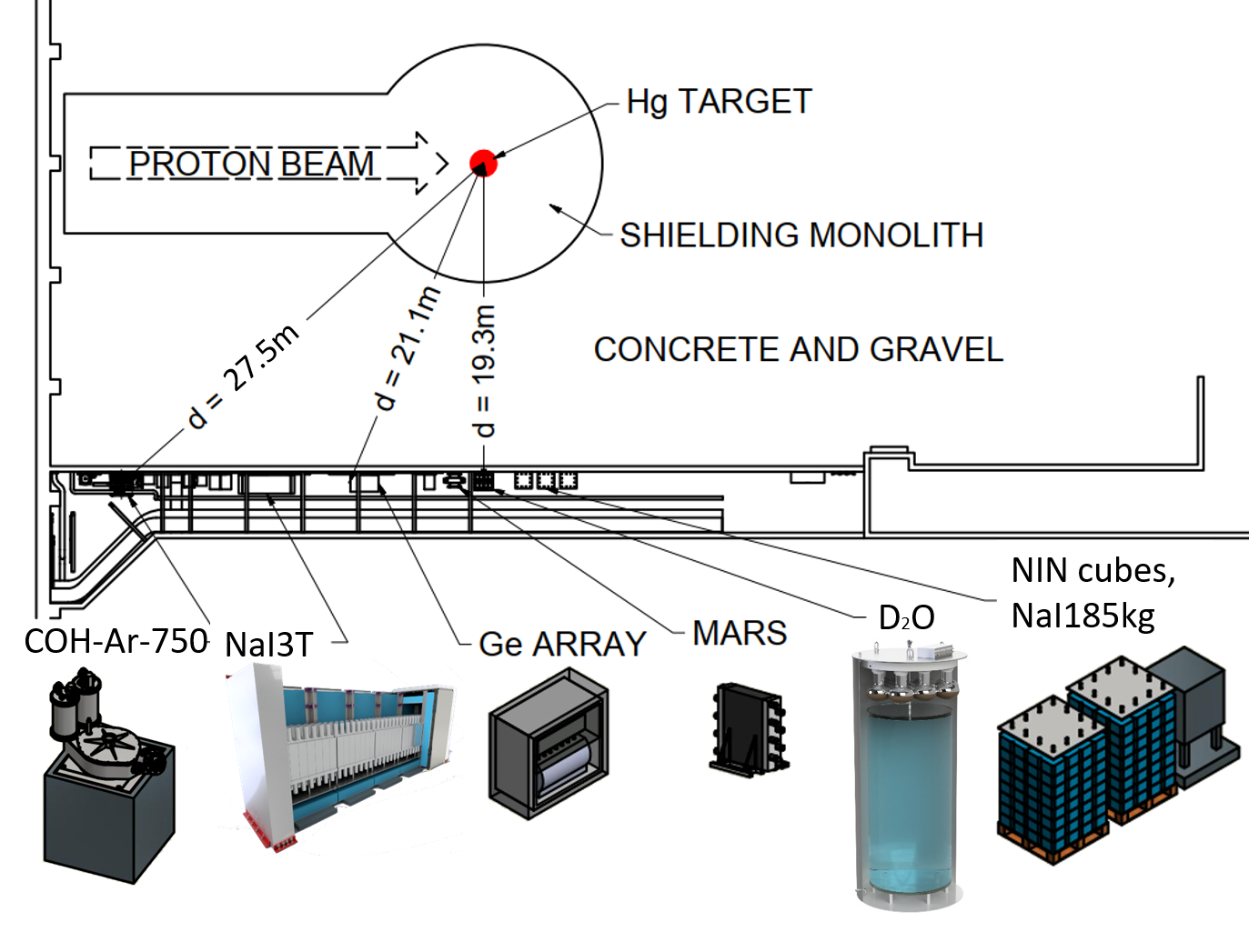}
\caption{Planned detector layout for the near future in Neutrino Alley. Figure from Ref.~\cite{COHERENT:2021xhx}.}
\label{fig:neutrino_corridor_current}
\end{figure}

\begin{figure}[h]
\includegraphics[width=3.5in]{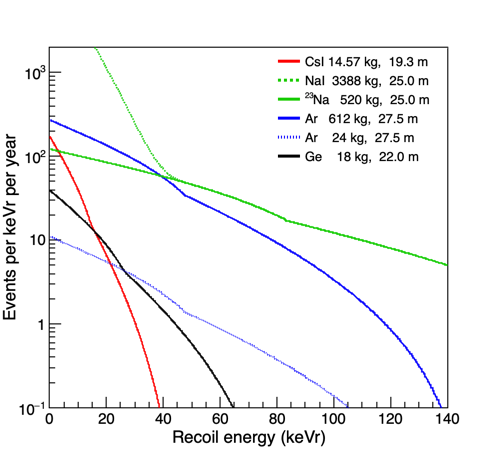}
\caption{Differential recoil spectra for COHERENT's CEvNS targets, scaled by mass and distance from the source.  No efficiencies or backgrounds are included.  The kinematics are manifest in that lighter targets have a lower yield overall, but more nuclei are kicked to high recoil energy.  }
\label{fig:diffspec}
\end{figure}

The COHERENT collaboration was created formally in 2014, following a workshop in 2012~\cite{Bolozdynya:2012xv}. COHERENT's primary goal is to pursue CEvNS measurements at the SNS. Further measurements including inelastic neutrino-nucleus interactions (see Sec.~\ref{sec:inelastics}) and accelerator-produced dark matter searches are additional goals. COHERENT has already deployed multiple detectors in Neutrino Alley and has at the time of this writing detected CEvNS in two target nuclei, with several additional deployments underway or planned.    
The CEvNS targets deployed by COHERENT span a range of neutron number values, with the aim of demonstrating the $N^2$ dependence of the cross section.   Table~\ref{tab:detectors} summarizes COHERENT's deployments.  
Figure~\ref{fig:neutrino_corridor_current} shows the layout of Neutrino Alley with COHERENT detectors.
Figure~\ref{fig:diffspec} shows the expected CEvNS recoil spectra for interactions in each detector type.

\begin{table*}[htbp]
	\centering
 	\begin{tabular}{c|c|c|c|c|c}
		\hline
 		Nuclear      & Detector           & Mass       & Distance from    & Dates       & Primary   \\

		target       & technology                       & (kg) & source (m) &   &  physics\\ 
		\hline 

	CsI[Na] & Scintillating crystal & 14.6       & 19.6             & 2015-2019   & CEvNS\\ 
 Pb, Fe & Liquid scintillator & 1000     & 19     &   2015-      
 & NINs\\
	NaI[Tl] & Scintillating crystal & 185    & 21               & 2016- & Inelastics\\

	  	LAr          & Noble scintillator          & 24         & 27.5             & 2017- & CEvNS \\
		LAr          & Noble scintillator          & 612        & 27.5             & proposed & CEvNS, inelastics\\

	D$_2$O & Cherenkov & 600 kg      &       22         & 2022- & Flux, inelastics\\
	
		Ge      & HPGe PPC              & 18      & 21                & 2022- & CEvNS\\

		NaI[Tl] & Scintillating crystal & 3388       & 24                & 2022- & CEvNS, inelastics\\
		
	CryoCsI & Scintillating crystal & TBD &  TBD & proposed & CEvNS \\			
 
		\hline
	\end{tabular}
	\caption{Parameters for the current and future COHERENT detector subsystems. 
    \label{tab:detectors}}
 \end{table*}

\subsection{Backgrounds}

Backgrounds for CEvNS measurements include anything that can masquerade as a signal from a nuclear recoil.   These backgrounds can be broken into two general categories:  steady-state backgrounds, and beam-related backgrounds.  The steady-state backgrounds are uncorrelated with the beam, and can include ambient and intrinsic radioactivity, cosmogenics, and detector noise.  Steady-state backgrounds are dependent on the detector nature and location. The pulsed nature of the SNS beam gives an enormous advantage for mitigation of this background category: not only can events outside of the beam window be rejected, but the out-of-time-window data can be used to precisely measure the steady-state background inside the beam time window.  Therefore, only fluctuations of the steady-state background will matter for physics measurements.  

\subsubsection{Beam-Related Neutrons}
Beam-related backgrounds are in principle more pernicious than steady-state backgrounds, as the off-beam measurement strategy cannot be used to reduce and constrain it.  The SNS produces neutrons, and fast neutrons can arrive in time with the beam pulse (although with a different time of flight delay with respect to the neutrinos).
Therefore it is critical not only to shield from this background as much as possible, but also to carefully model and characterize it using measurements and simulation.

Before the first deployment, the COHERENT collaboration spent
about two years measuring backgrounds in several locations around the SNS using the Sandia Scatter Camera~\cite{Brennan:2009} and SciBath~\cite{Tayloe:2006ct} detectors.  Neutrino Alley turned out to have very low background, thanks to a thick layer of material between the neutrino source and the corridor.  An ongoing neutron monitoring program including the MARS detector~\cite{Roecker:2016juf} and additional scintillators continues to provide information for background modeling.

\subsubsection{Neutrino-Induced Neutrons}\label{sec:nins}
One particularly interesting and difficult-to-shield background is from neutrino-induced neutron (NIN) reactions on shielding materials surrounding the CEvNS detectors.  Neutrinos at stopped-pion energies are capable of interacting in high-Z shielding components, such as lead, to eject few-MeV neutrons via CC (and subdominantly NC) interactions~\cite{Kolbe:2000np,Vaananen:2011bf}.  These neutrons can then make their way into the sensitive CEvNS target to create recoils that mimic CEvNS in time with the neutrino flux.  The NIN cross section is smaller than the CEvNS one, but is nevertheless large enough to have the potential to create a non-negligible background, unless shielding is designed carefully to avoid large quantities of lead close to the CEvNS target. NINs can be abated with designs that moderate them with a few tens of cm of water or polyethylene inside the last layer of lead shielding.

The NIN cross section has a large theoretical uncertainty, and is itself of interest, as this kind of interaction is useful for supernova detection in detectors such as HALO~\cite{Duba:2008zz} and for understanding of nucleosynthesis~\cite{qian1997:NINSnucleosynthesis, woosley1990:nuProcess}.
COHERENT has deployed two dedicated NIN detectors to study this interaction in systems dubbed the ``neutrino cubes,''  one with lead and another with iron. The neutrino cubes each consist of four cells of organic liquid scintillator for the detection of neutrons. For neutron-proton interactions that deposit sufficient energy ($\sim$50 keV), pulse shape discrimination algorithms performed on the recorded scintillation waveform separate neutron interactions from background gammas. The cells are surrounded by the Pb and Fe neutrino targets, which are themselves hermetically sealed in a veto to reject neutrons produced by muons traversing the Pb or Fe, and $\sim$22 cm of water to moderate any background neutrons present in Neutrino Alley.  First results are expected in 2022.

\section{COHERENT RESULTS}\label{sec:results}

COHERENT made the first measurement of CEvNS in 2017 using its CsI detector.  This was followed in early 2020 with a second measurement on argon.
We can note that these are the only instances so far of a low-energy NC neutrino-hadron 
interaction with event-by-event spectral information.

\subsection{First CEvNS Measurement on CsI}

The first COHERENT CEvNS measurement~\cite{Akimov:2017ade} was made using a CsI[Na] crystal detector~\cite{Collar:2014lya} of 14.57 kg mass and viewed by a single photomultiplier tube (PMT).  The detector was deployed 19.6 m from the SNS target in Neutrino Alley.  Shielding included (from inside to outside): 3 inches of high-density polyethylene, 2 inches of low-background lead, 4 inches of regular lead, 2 inches of muon veto panels and 4 inches of water.  
Cesium iodide is a high-light-yield scintillator, which can be robustly operated at room temperature.
Sodium-doped CsI was chosen due to favorable afterglow properties with respect to thallium-doped CsI.  A 2-kg test crystal was calibrated at the University of Chicago~\cite{Scholz:2017ldm}, with 13.35 photoelectrons  (PE) keVee$^{-1}$ light yield measured. This amounts to about 1.2 PE per keVr. The QF for a similar crystal was measured at the Triangle Universities Nuclear Laboratory (TUNL) using neutrons. For the first measurement, the QF was estimated as 8.78\%, and assumed to be flat, with uncertainty estimated as the range between two measurements.  The QF uncertainty was the dominant uncertainty on the first measurement.

\begin{figure}[h]
\includegraphics[width=3in]{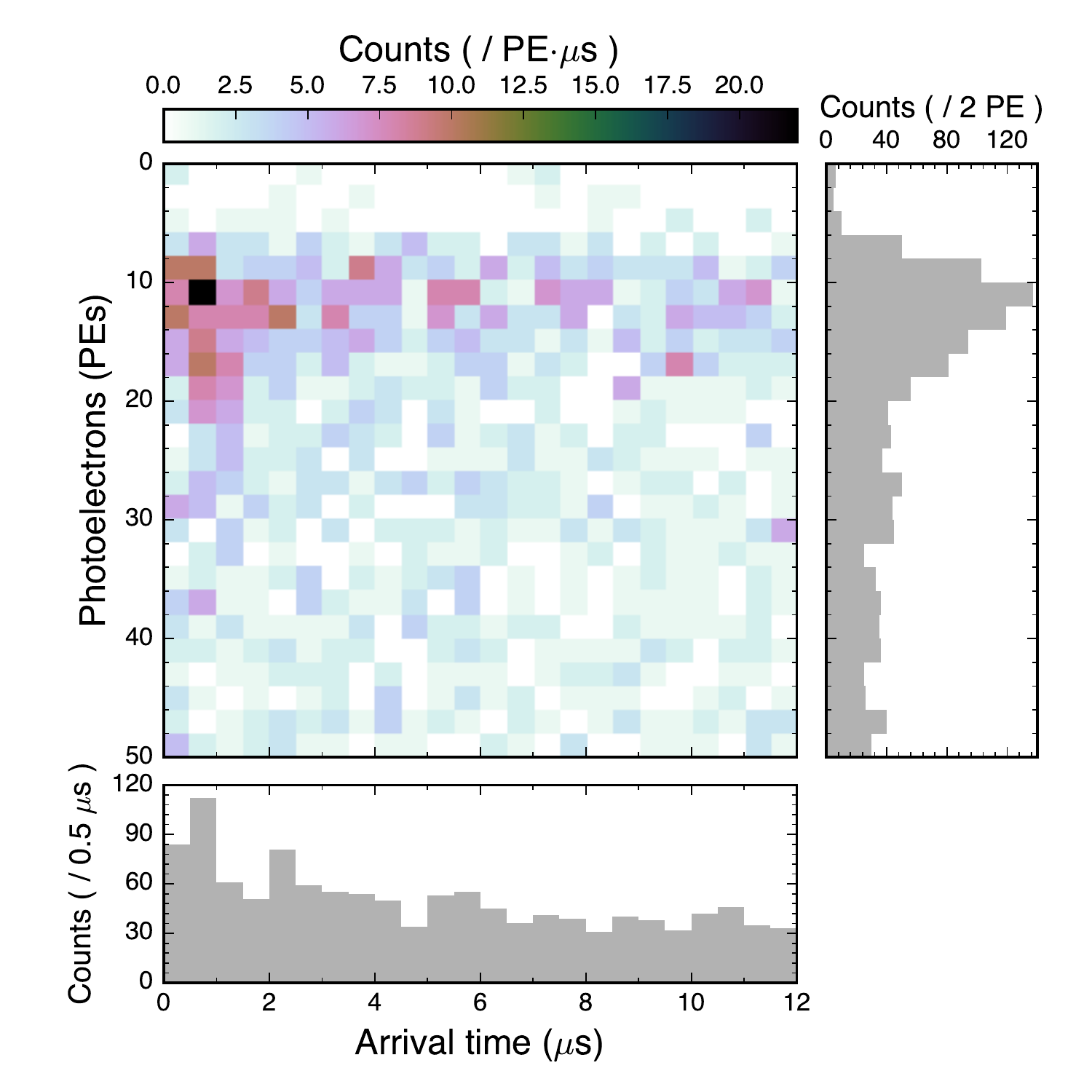}
\caption{Two-dimensional PE vs time plot for beam-on CsI data for the first-result sample~\cite{Akimov:2017ade}. Plot from Ref.~\cite{Akimov:2018vzs}.}
\label{fig:2dfirstcsi}
\end{figure}

Data-taking corresponded to $\sim 1.76 \times 10^{23}$ protons on target (POT), for a 7.45~GW-hr exposure at the SNS.   
Quality cuts to the raw data, described in detail in Ref.~\cite{Scholz:2017ldm}, were applied to remove coincidences with the muon veto, dead time from PMT saturation, and digitizer overflow.   Cuts were applied to reject  phosphorescence-related afterglow hits, and to remove accidental coincidences between the Cherenkov emission in the PMT window and dark counts. Rise time cuts were applied to select good pulse shapes.  The overall efficiency for event selection as a function of recoil energy turned on at about 6.5~pe and reached a maximum plateau of about 66\%  at around 16 PE.   
Two analyses with slightly different cut optimizations yielded consistent results.  
The optimized time window was 6-30 PE in 0-6000 microseconds from beam start.  The public release of this data set can be found in Ref.~\cite{Akimov:2018vzs}.  Figure~\ref{fig:2dfirstcsi} shows the two-dimensional PE and time distribution of the background-subtracted data.

\begin{figure}[h]
  \begin{minipage}[c]{0.5\textwidth}
\includegraphics[width=1.8in]{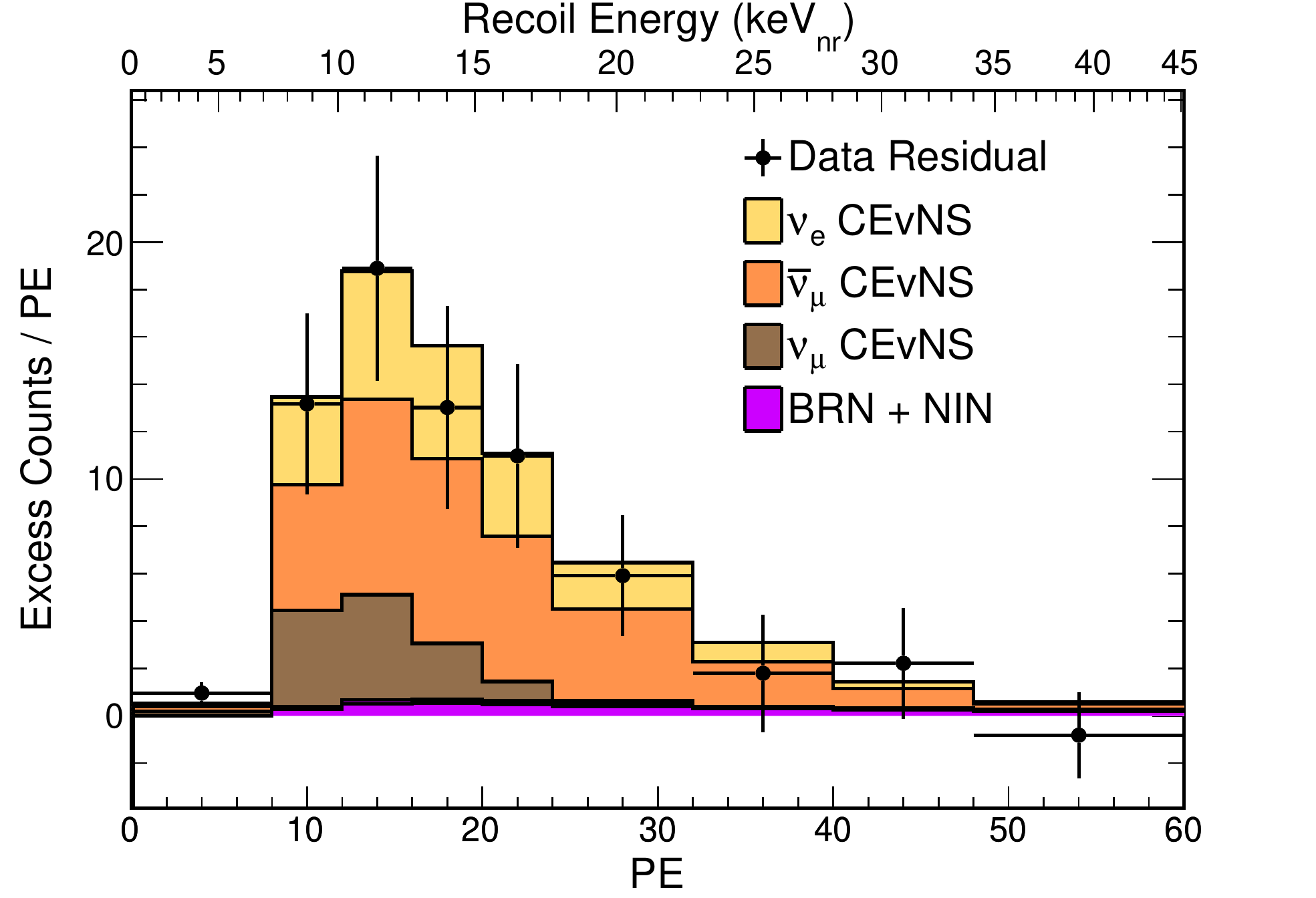}
\end{minipage}\hfill
  \begin{minipage}[c]{0.5\textwidth}
\includegraphics[width=1.8in]{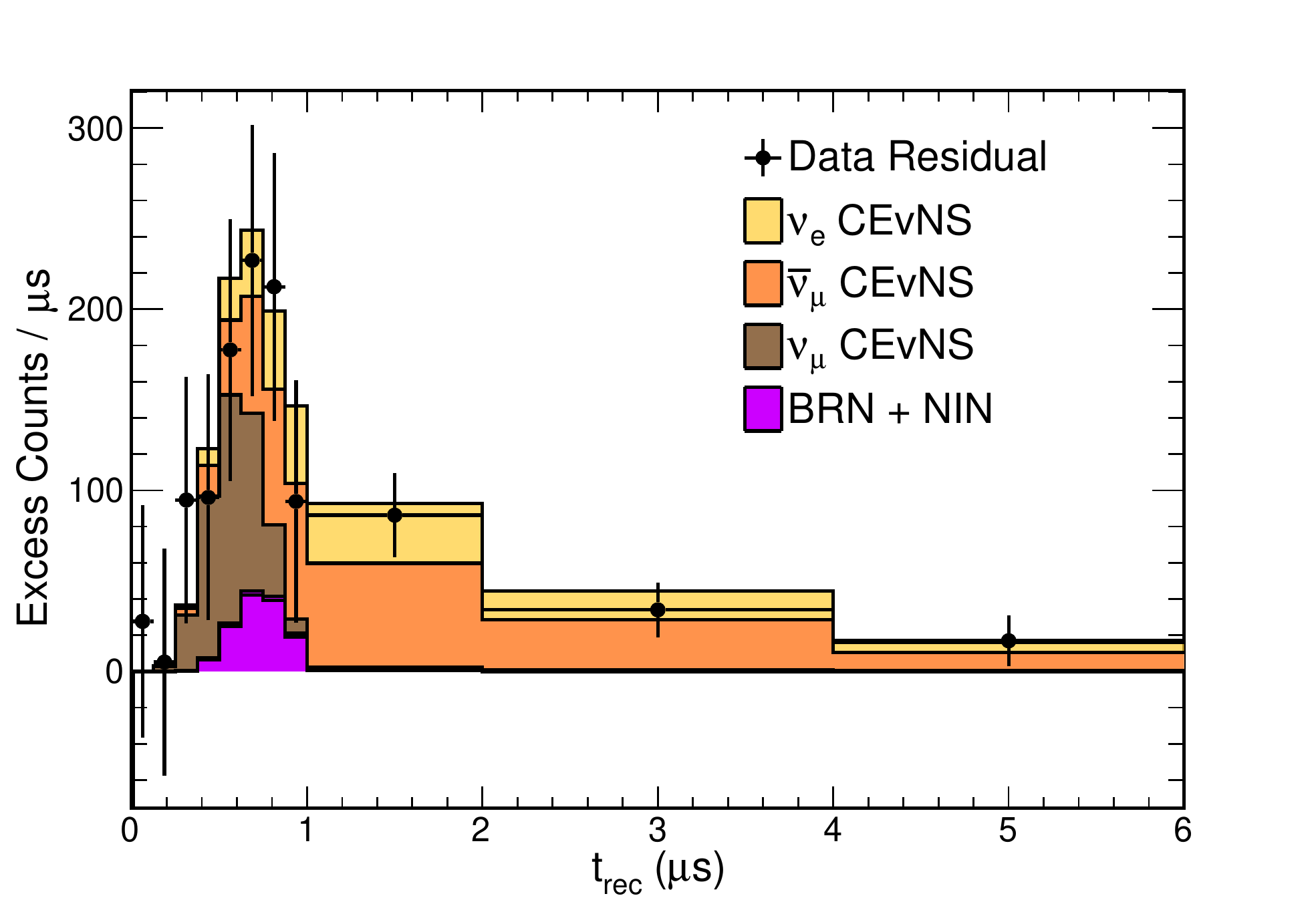}
\end{minipage}\hfill
\caption{Steady-state-background subtracted PE and time distributions from Ref.~\cite{Akimov:2021dab} for the full CsI dataset. The yellow, orange and brown regions represent the flavor content of the SM CEvNS expectation, and the magenta regions represent the estimated background neutrons (BRN: beam-related neutrons; NIN: neutrino-induced neutrons).}
\label{fig:fullcsi}
\end{figure}

Steady-state backgrounds were evaluated with an anticoincidence time window opened before each trigger.
Beam-related neutron backgrounds were evaluated using an EJ-301 liquid scintillator cell which was deployed inside the shielding before the CsI crystal deployment. Simulations showed a result consistent with expectation.  A total of 11 beam-related neutron events were expected (the fits for the first measurement ignored the NIN signals).

A  likelihood analysis taking into account the signal and background shapes in time and PE yielded a result of $134 \pm 22 $ CEvNS events, with the uncertainty being primarily statistical~\cite{Rich:2017lzd}. The SM prediction for this analysis is  $178 \pm 43 $ CEvNS events.  The observed event rate was consistent with the SM prediction within uncertainties.

The systematic uncertainties in the CsI[Na] paper predicted counts were estimated to be:  25.5\% due to quenching factor uncertainties, 5\% due to form factor uncertainties, and 5\% due to uncertainties on efficiency estimate.  The systematic uncertainties on beam-on background were estimated to be 25\% (note that systematic uncertainties on steady-state background can be reduced to effectively zero, because these backgrounds can be measured off beam pulse.)   The overall uncertainty on the flux was estimated to be 10\%.  

The CsI detector was decommissioned and removed from Neutrino Alley in July 2019. 

Analysis from the full CsI data sample with additional data amounting to more than twice the statistics and improved QF evaluation~\cite{Akimov:2021ggw} is now complete~\cite{Akimov:2021dab}.  Data selection cuts and efficiencies are improved with respect to the initial analysis.  This dataset corresponds to 3.2$\times 10^{23}$ POT and a 13.99 GW-hr exposure.  Figure~\ref{fig:fullcsi} shows the PE and time distributions along with signal and background predictions.  The SM expectation for the number of observed events is 341$\pm$43, and the best fit to the data resulted in 306 $\pm$ 20 events, $17.3 \pm 4.5$ beam-related neutron events (all in the prompt window) and $5.5\pm 2.0$ NINs.
Total systematic error on the prediction due to QF is reduced to about 4\% for this analysis.  The dominant systematic uncertainty is now the 10\% uncertainty on the neutrino flux normalization.

\subsection{First CEvNS Measurement on Argon}

The next detector to be deployed in COHERENT's suite was a 24-kg single-phase argon detector, dubbed COH-Ar-10~\cite{Akimov:2019rhz} (formerly CENNS-10), in which the CEvNS recoil energy is detected via wavelength-shifted 128-nm scintillation photons.  Argon also has the capability for pulse shape discrimination that can be used to distinguish nuclear recoils from electronic (electron or gamma) events, which reduces background. 

The COH-Ar-10 detector is located 27.5 m from the source.  The detector was borrowed from Fermilab~\cite{Brice:2013fwa} and refurbished at Indiana University.  The argon detector vessel has a cylindrical volume of 21 cm diameter and 61~cm height, and it is lined with Teflon$^{\rm{TM}}$ coated with the wavelength-shifter tetraphenyl butadiene (TPB).  The argon volume is viewed by two 8-inch PMTs.  Layers of shielding outside the steel detector vessel and cryostat include (from inside to outside) 23~cm of water, 1.2~cm of copper and 10~cm of lead.
This detector was first installed in 2016 and then upgraded in spring of 2017 with additional TPB coating on the PMT and inner detector vessel.  The light yield is determined to be 4.5 pe/MeV, which corresponds to a recoil threshold of around 20 keVr.   Calibrations have been done with $^{83m}$Kr~\cite{Akimov:2020xsa} injected inside the detector volume and with external Am/Be, $^{57}$Co and $^{241}$Am sources.  Beam-related neutron backgrounds were carefully studied using different detector shielding configurations and sidebands~\cite{Akimov:2019rhz}. 

\begin{figure}[h]
\includegraphics[width=3in]{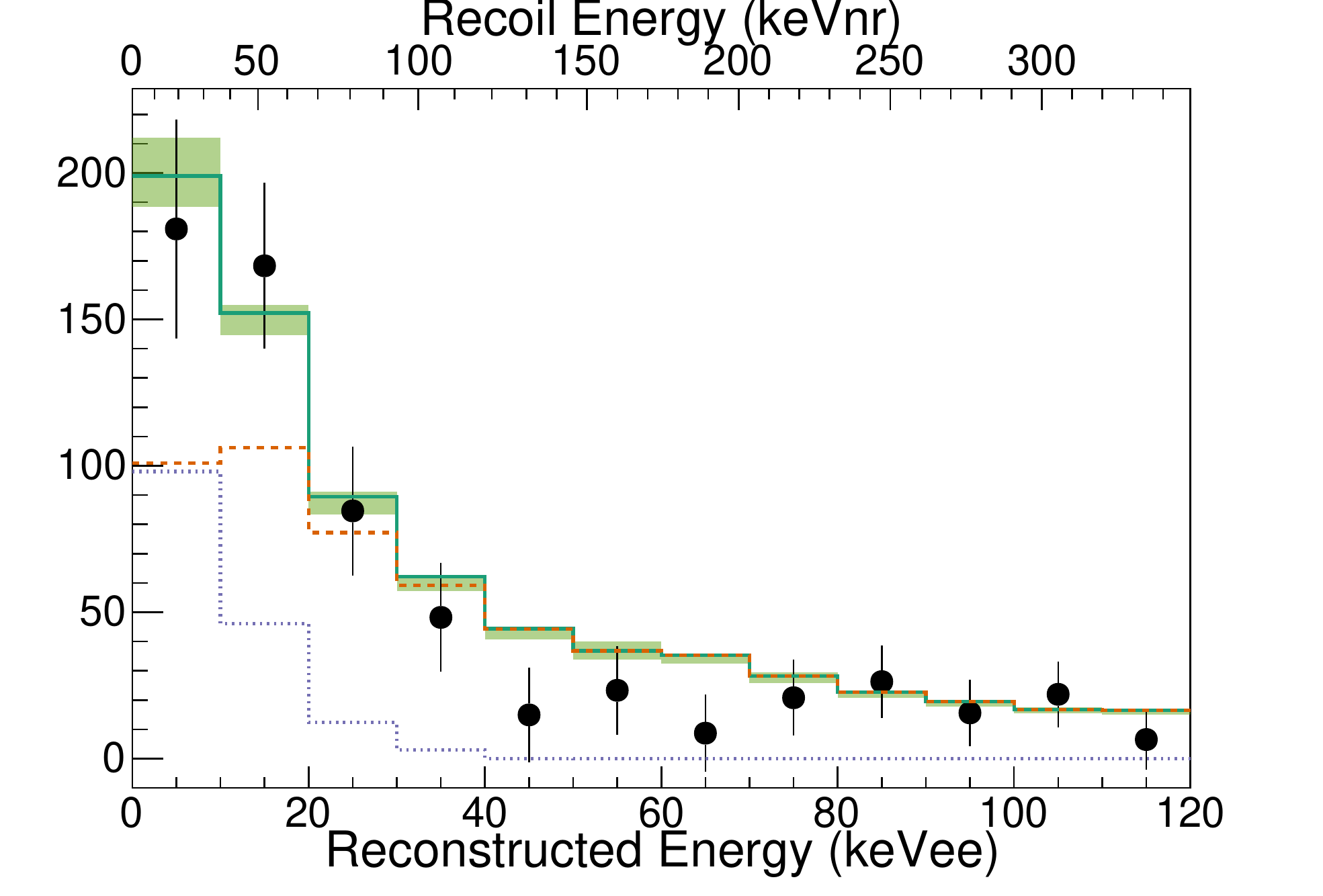}
\caption{Argon recoil spectrum with best-fit result.  The points are the data.  The blue dotted line is the fit CEvNS contribution, the red dashed line is the fit beam-related neutron contribution, and the solid line is the total. The green band shows 1$\sigma$ systematic uncertainty.  Figure from Ref.~\cite{COHERENT:2020iec}.}
\label{fig:argonresult}
\end{figure}

The first CEvNS measurement in argon was made using a 6.12 GW-hr, 13.7$\times 10^{22}$ POT data sample and extracted from a likelihood fit to time, recoil energy and pulse-shape-discrimination parameter
~\cite{COHERENT:2020iec,Akimov:2020czh}. Two independently developed analyses gave consistent results.   The recoil energy distribution for one of these is shown in Fig.~\ref{fig:argonresult}.  
Compared to the CsI result, QF uncertainty is small; however beam-related neutron backgrounds at the argon detector location are larger.  The largest systematic uncertainties are flux normalization and detector-related uncertainties. The public release of the data can be found in Ref.~\cite{Akimov:2020czh}.

\subsection{Physics Interpretation of COHERENT Results}

\subsubsection{Interpretation of CEvNS Measurements}

\begin{figure}[h]
\includegraphics[width=4in]{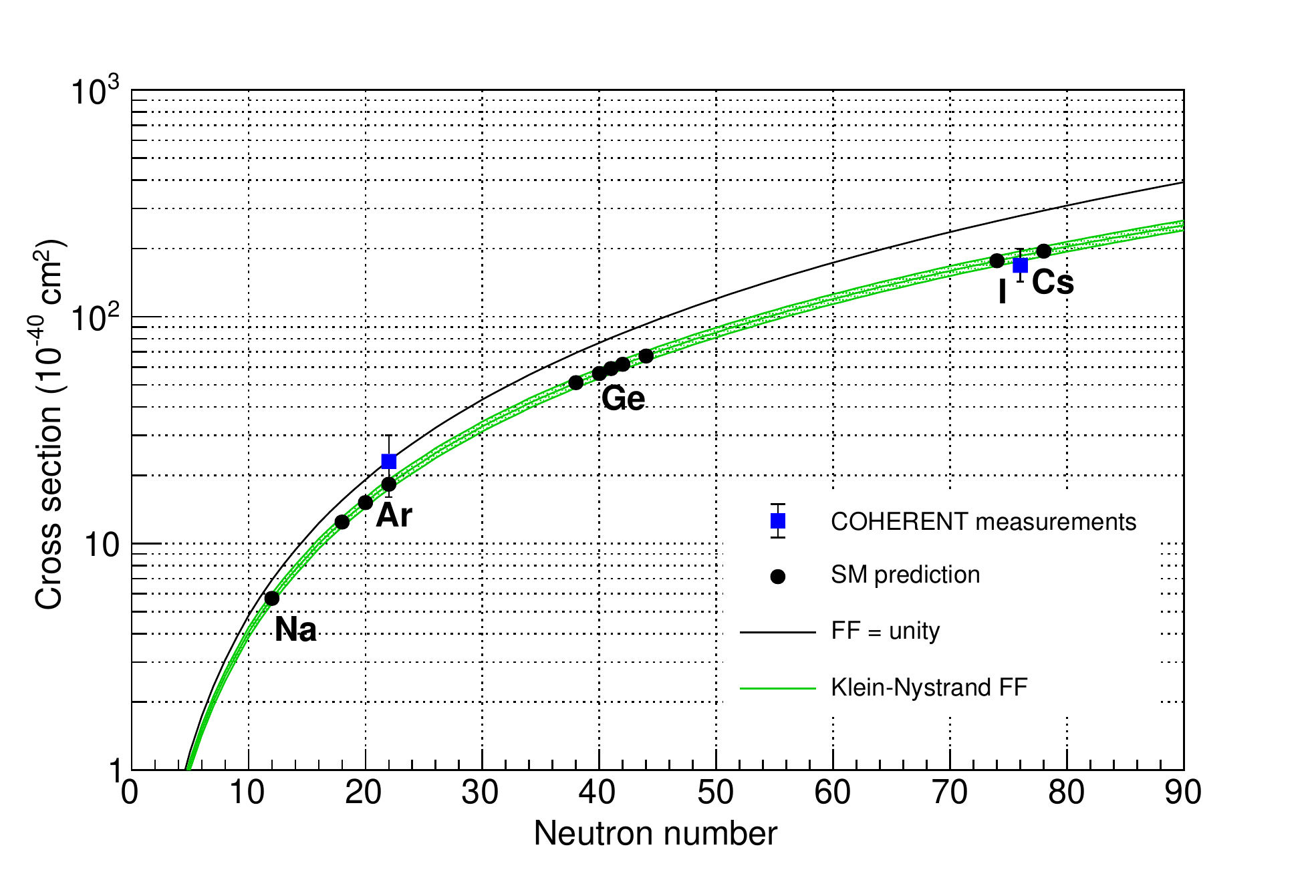}
\caption{Cross section averaged over a stopped-pion spectrum as a function of neutron number $N$ in the target.  The thin black line is for unity form factor; the green line shows the SM prediction with thickness corresponding to $\pm$3\% uncertainty on the nuclear radius in the form factor.  The points with error bars are the COHERENT measurements on argon and full-dataset CsI.}
\label{fig:Nsq}
\end{figure}

\begin{figure}[h]
\includegraphics[width=3in]{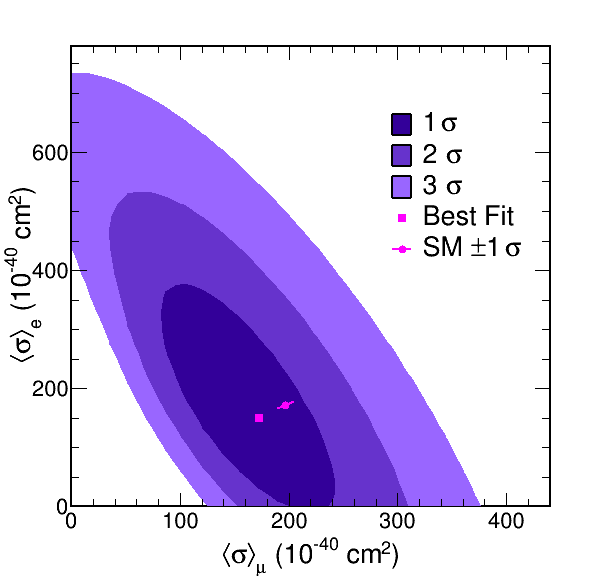}
\caption{Flux-averaged cross sections for muon and electron flavor separately, from Ref.~\cite{Akimov:2021dab}.}
\label{fig:flavxscn}
\end{figure}

The two COHERENT measurements so far are shown as total (all-flavor) cross sections averaged over the stopped-pion flux in Fig.~\ref{fig:Nsq}.  Both are consistent within 1$\sigma$ with the SM predictions.  From the full CsI dataset,  $\sin^2\theta = 0.220^{+0.028}_{-0.026}$ is inferred.
For this CsI data set, the inferred \textit{flavor-dependent} cross sections, determined using timing, for muon and electron flavor separately are shown in Fig.~\ref{fig:flavxscn}.    

\begin{figure}[h]

  \begin{minipage}[c]{0.5\textwidth}
\includegraphics[width=2in]{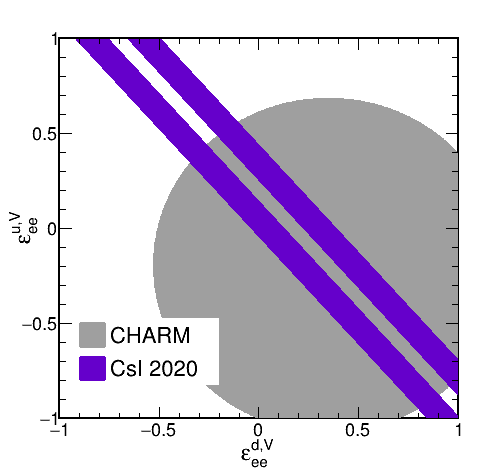}
\end{minipage}\hfill
  \begin{minipage}[c]{0.5\textwidth}
\includegraphics[width=2.2in]{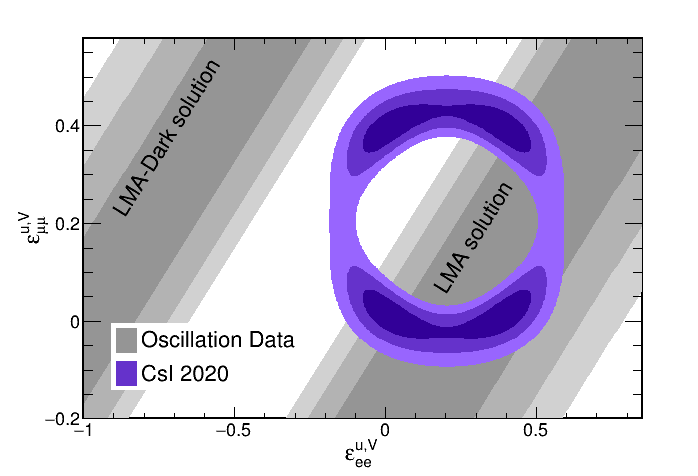}
\end{minipage}\hfill

\caption{Example constraints at 90\% C.L. on NSI parameters using the full CsI data, from Ref.~\cite{Akimov:2021dab}.  In each case, the other NSI parameters are assumed zero.  The parameters in the right-hand plot are relevant for understanding the ``LMA-dark" solar neutrino oscillation parameters. }
\label{fig:NSIresult}
\end{figure}

These first results are sufficient to make some meaningful physics constraints on BSM neutrino physics.    Fig.~\ref{fig:NSIresult} shows example 90\% allowed regions for some  of the  $\varepsilon$ parameters described in Sec.~\ref{sec:nsi}.   Examples of additional studies that use COHERENT data to constrain NSI can be found in Refs.~\cite{Coloma:2017ncl,Denton:2018xmq,Abdullah:2018ykz,Coloma:2019mbs,Canas:2019fjw, Papoulias:2019xaw}.

\subsubsection{Dark Matter Search Results}

\begin{figure}[h]

\includegraphics[width=3in]{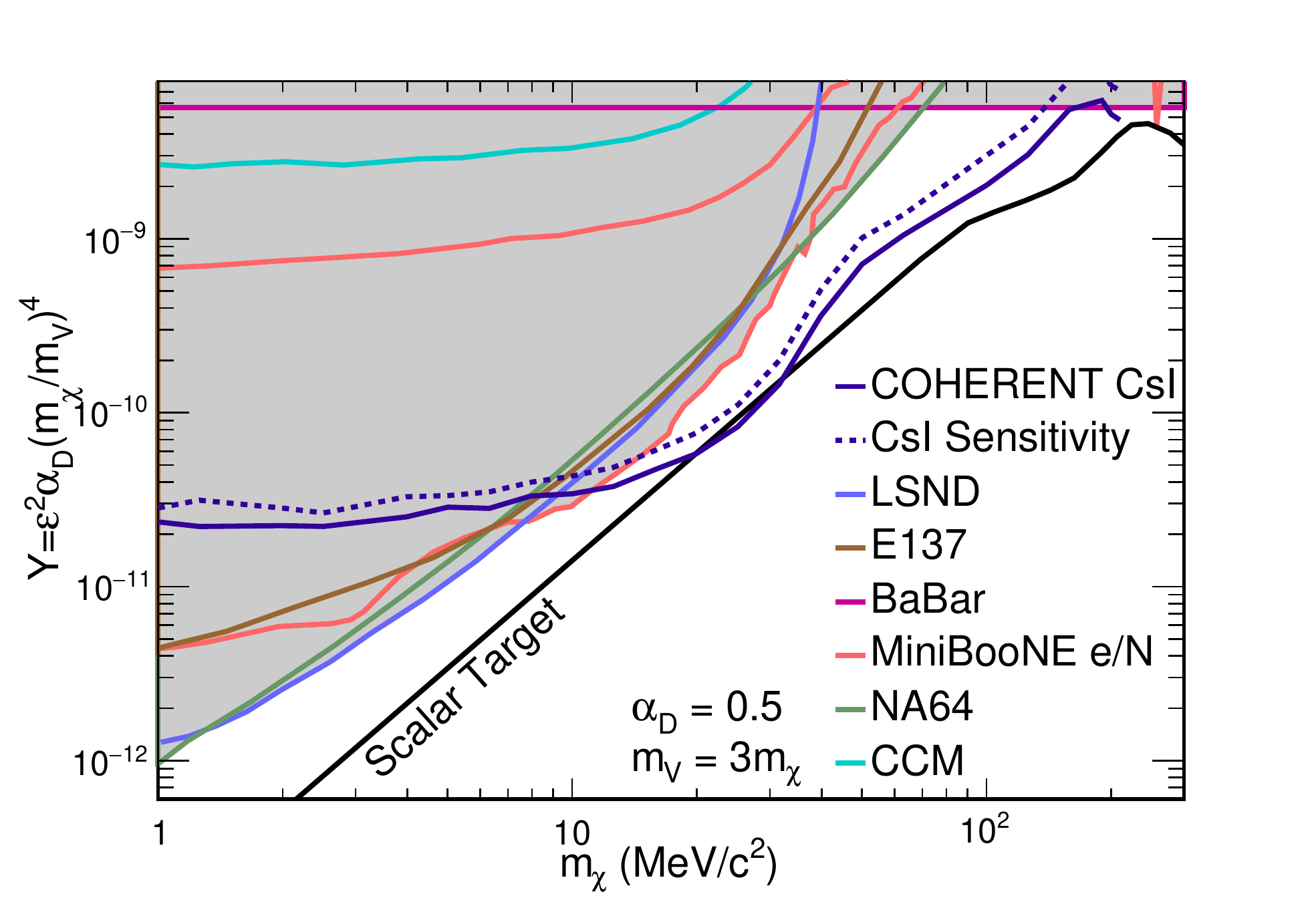}

\caption{Constraints at 90\% C.L. from the full COHERENT CsI dataset in coupling-mass parameter space for scalar dark matter produced at the SNS under the assumption of a benchmark vector portal model.    The parameter $Y$ is defined as $Y=\varepsilon^2 \alpha_D \left( \frac{m_\chi}{m_V}\right)^2$, where $m_\chi$ and $m_V$ are masses of a scalar DM particle $\chi$ and vector portal particle $V$, respectively; $\varepsilon$ quantifies kinetic-mixing coupling of $V$ with a SM photon; and $\alpha_D$ quantifies coupling in the $V\rightarrow \chi \bar{\chi}$ decay.   Conservatively chosen values of $\alpha_D$ and $m_V$ are indicated. Shown also are limits from other experiments, including CCM~\cite{CCM:2021leg}, which also uses a nuclear recoil signature and a stopped-pion-type beam. Figure from Ref.~\cite{Akimov:2021yeu}. }
\label{fig:dmresult}
\end{figure}

Furthermore, COHERENT CsI data have been used to set constraints on sub-GeV dark matter.  Fig.~\ref{fig:dmresult} shows the constraint from Ref.~\cite{Akimov:2021yeu} based on non-observation of an excess of prompt counts with the expected recoil spectrum for a DM model, for a specific model of scalar dark matter.  This is the first constraint for a  model of this type that goes beyond the cosmological expectation in the $\sim$25 MeV/$c^2$ mass range.  A range of other models can be similarly constrained~\cite{Akimov:2021yeu}.

Joint physics analyses with COHERENT Ar and CsI datasets are currently in progress.

\section{THE FUTURE COHERENT PROGRAM}

The COHERENT collaboration  is currently pursuing several additional detector technologies for CEvNS, to span a range of $N$ values, as well as detectors to address additional physics goals.  

\subsection{Germanium}

The COHERENT collaboration is preparing to deploy 18 kg of p-type point contact (PPC) Ge detectors in early 2022.  Germanium offers a CEvNS target with an intermediate $N$ value. The favorable features of this type of detector are superior energy resolution, low energy threshold  ($\sim 1$~keVr), and low background.  Signal timing is not as fast as that for scintillation detectors, but should be sufficient to exploit the SNS beam window for background rejection.  Previous deployments of similar technology include
    CoGeNT  \cite{Aalseth:2008, Aalseth:2011, Aalseth:2011wp, Aalseth:2012if}, \textsc{Majorana Demonstrator}~\cite{Abgrall:2014}, GERDA~\cite{Gerda:2017}, LEGEND-200~\cite{Abgrall:2017syy}, CDEX~\cite{Zhao:2016dak}, TEXONO~\cite{lin2009:texono}, and CONUS~\cite{Lindner:2017,Hakenmuller:2017} (the latter having recently been used for CEvNS searches at a reactor). A multiport dewar will accommodate an array of eight $\sim$2-kg detectors from Mirion Technologies.  These will be shielded by copper, polyethylene and lead, and surrounded by a plastic scintillator muon veto.   Approximately 600 interactions per SNS-year are expected. 
 
\subsection{Sodium Iodide} \label{sec:nai}   
    
The COHERENT collaboration is deploying several tonnes of NaI[Tl] scintillator to Neutrino Alley,  
for a CEvNS measurement on the lightest nuclei so far considered.

Due to the small nuclear size, the impact of any uncertainty on the nuclear form factor is minimized, improving the possible BSM reach of this measurement. The lighter Na nuclei will result in a larger fraction of signal recoils above threshold and less uncertainty due to the threshold (see Figure \ref{fig:diffspec}); however, they will also suffer from a dramatically reduced cross section due to relatively small number of neutrons contributing to their weak nuclear charge. In addition, the majority of the mass of a NaI[Tl] scintillator is composed of iodine nuclei, necessitating a larger detector deployment. The collaboration currently has deployed a 185-kg prototype detector array composed of 24 7.7-kg crystals. A much larger three-tonne array will be operational in early 2022.

Sodium is not a pure even-even nucleus. The unpaired proton spins give rise to axial currents which are not as well understood, and could complicate any search for non-standard neutrino interactions. The impact is larger for light nuclei, where it has not been washed out by the coherent vector coupling to the neutrons. While a small effect ($<$5\%), the axial-current contribution does impact the shape of the recoil spectra at the higher energies, which is easier to measure with the higher energy, lighter-mass recoils.

The large mass of iodine nuclei in the NaI[Tl] scintillators serves as a convenient target to measure the CC electron neutrino interaction on $^{127}$I. The CC signal from the resulting electron has orders of magnitude more observable energy (tens of MeV) than the CEvNS recoil and is also easy to contain in a the large array of crystals available. The low-energy neutrino CC cross section for $^{127}$I($\nu_e$, e$^-$)$^{127}$Xe is a sensitive measure of the quenching of the axial current g$_A$. The current $\sim$40\% theoretical uncertainty is almost entirely due to the unknown magnitude of g$_A$ quenching~\cite{PhysRevC.50.1702}. A previous measurement of this exclusive process exists \cite{Distel:2002ch}; however, the uncertainty ($\sim$33\%) is insufficient to address these theoretical questions. A parallel measurement could have important implications for ab initio calculations of neutrino-nucleus interactions and nuclear matrix elements in neutrinoless double beta decay. 
    
\subsection{Tonne-scale Argon}    
    
The 24-kg COHERENT argon detector finished its first production run of data-taking in fall of 2021 and will be used to explore detector options for a future liquid argon deployment in its current location in the near term.
    
The COHERENT collaboration plan is to scale up the argon detector in the same location to approximately 610 kg of fiducial mass (total 750 kg), shielded with $\sim$10~cm of lead and $\sim$15~cm of water, and concrete.  The argon cooling will be via both liquid nitrogen and pulse-tube cryocoolers.   Designs for light sensors under consideration include either 3-inch PMTs coated with TPB or VUV-sensitive silicon photomultipliers, or both.   Xenon doping of the argon is also under consideration.
An upgraded detector will provide much expanded statistics and potentially higher light yield and lower threshold, as well as good sensitivity to inelastics (see Sec.~\ref{sec:inelastics}). 

The dominant steady-state background for COH-Ar-10 is $^{39}$Ar decays, which has a beta spectrum with endpoint at 0.565~MeV. There is promising potential to reduce this background with argon depleted in this cosmogenic isotope from underground sources~\cite{Xu:2015hja,Alexander:2019uvv}.

\subsection{Other Future COHERENT CEvNS Detectors}

Several other possibilities have been considered for Neutrino Alley and beyond at the STS, although none of the following have yet been adopted as a formal collaboration plan.  Both nuclear recoil and inelastic interaction physics can be addressed, as well as BSM particle searches.

A possibility under consideration is cryogenic scintillator.  Undoped inorganic scintillators such as CsI and NaI  operated at cryogenic temperatures have been shown to have high light yield ($\sim 30$ photoelectrons per keVee~\cite{Chernyak:2020lhu,Ding:2020uxu}, approximately double the light yield of the CsI[Na] crystal used previously by COHERENT), and can potentially achieve low recoil threshold.  Such detectors can be deployed at the tens of kg scale.

 There are many other possibilities, including additional targets (e.g., neon, xenon) to cover a wider range of $N$ values for CEvNS, liquid argon time projection chambers, different detector configurations (HALO-like detector for NINs), and new technologies (e.g., bubble detectors~\cite{Kozynets:2018dfo}).   Detectors sensitive to the directionality of recoils (e.g.~\cite{Vahsen:2020pzb}) can in principle exploit this directionality for selection of signal against background, and for discrimination of BSM models~\cite{Abdullah:2020iiv}.  However, many such directional detectors make use of gas phase and therefore require large volume, which makes siting a challenge.

\subsection{Reducing Signal Uncertainties}

For future COHERENT deployments,  statistical uncertainties will shrink substantially, while systematic uncertainties may vary by detector. 

\subsubsection{Reducing Flux Uncertainties with Heavy Water Detector}\label{sec:d2o_det}
A systematic uncertainty shared by all detectors in COHERENT's suite is the neutrino flux uncertainty.  Thanks to the high fraction of pions that decay at rest, neutrino spectral shape uncertainty is very small.  There is a small fraction of pions that decay in flight, but due to the SNS energy and dense mercury target the production rate of these is below 1\% relative to the decay-at-rest rates and hence these contribute negligibly to the overall neutrino interaction rates. However, the flux normalization uncertainty is evaluated to be 10\% by comparing the output of several pion-production models in the beam simulation, and validated by comparing to available data~\cite{Akimov:2021geg}.  This normalization depends on pion production and interaction cross sections with the SNS target materials, and modeling can potentially be improved using  measurements of pion production for protons incident on a similar large nucleus (as was done for the T2K experiment with NA61/SHINE data~\cite{Berns:2018tap}.) 
Measurements using a proposed new low-energy beamline at CERN may help reduce this uncertainty~\cite{lowebeamline}.

However, a more direct way of reducing flux normalization uncertainty is to measure the $\nu_e$ flux. COHERENT will deploy a heavy water detector with this aim.
The neutrino cross section on deuterons is well understood theoretically; see, e.g.,~\cite{Acharya:2019fij}.
The COHERENT D$_2$O detector will be deployed in Neutrino Alley, approximately 20~m from the SNS target.  COHERENT is aiming to deploy two modules containing 600~kg each of heavy water~\cite{COHERENT:2021xhx}. Each module has a clear cylindrical acrylic vessel 70~cm in diameter and 140~cm tall containing the heavy water. Within this well-defined volume, electrons from the $\nu_e$-d CC breakup reaction, $\nu_e+d \rightarrow p + p + e^-$, will produce Cherenkov radiation. Outside of the acrylic container, an outer volume of H$_2$O (the ``tail catcher''), with a thickness of 10 cm will be contained within a steel tank. Electrons that escape the central volume will still produce Cherenkov light in this tail catcher region, allowing a complete
integration of the total electron energy. Twelve 8'' PMTs, immersed in the H$_2$O, view the fiducial volume from above. The inner walls of the steel tank  will be  covered in reflective Tyvek. With such a reflective material, light collection within the D$_2$O volume remains relatively uniform. According to simulation, about 15~PE per MeV of energy deposition should be detected. The detector will be calibrated primarily by using Michel electrons from stopped cosmic muons, for which the energy spectrum matches the expected $\nu_e d$ spectrum well. Outside the steel tank that supports the PMTs and encloses the tail catcher, 5.08 cm of lead shielding and two layers of 2.54-cm plastic scintillator panels mitigate external backgrounds due to beam-related neutrons, radioactivity in the hall, and cosmic rays. The dimensions of the detector are driven by a space limitations in Neutrino Alley. In two years of data taking we expect to have $\sim$1200 $\nu_e d$ interactions, which will allow calibration of SNS neutrino flux normalization with 3\% accuracy. In addition to interactions  on deuterium, we will see about 170 CC reactions on $^{16}$O. Electrons from this reaction will have energies shifted down from neutrino interactions on deuterons by $\sim$15 MeV. Therefore, oxygen events produce negligible background for SNS calibration but will have value in themselves as a measurement on oxygen for the energy range of interest for supernova neutrino detection by water Cherenkov detectors like Super-Kamiokande~\cite{Abe:2016waf} and Hyper-Kamiokande~\cite{Abe:2018uyc}.

\subsubsection{Quenching Factor Measurements}\label{sec:qf}
For many of the COHERENT detector systems, the largest source of uncertainty when measuring the CEvNS rate is the determination of the fraction of nuclear recoils detectable above threshold. The majority of the CEvNS recoils occur near the detector threshold because the low-momentum transfer required for coherence results in a low-energy recoil. For nearly all of the systems, the detector threshold limitation reflects cutting-edge technological development often pioneered by the dark matter direct-detection community, in search of low-energy recoils from low-mass WIMPS. Understanding of these thresholds is paramount, but a major complicating factor is that nuclear recoils often produce  only a fraction of the corresponding energy deposition from an electronic interaction (electron or photon). This fraction, typically referred to as the quenching factor, can be of the order of 10-20\%, which is the case for the NaI[Tl] \cite{ GERBIER1999287, SIMON2003643, Stiegler:2017kjw, SPOONER1994156, TOVEY1998150, PhysRevC.92.015807}, CsI[Na] \cite{Akimov:2017ade}, HPGe \cite{Barbeau_2007, PhysRevLett.106.131302, PhysRevLett.21.1430, PhysRev.143.588, PhysRevLett.69.3425} and liquid argon detectors~\cite{COHERENT:2020iec} described here. 

Quenching factors are often not well understood enough to perform precision measurements of the CEvNS cross section. In general, it is the case that more robust QF measurements are required for all CEvNS detector systems. 
Such measurements are typically performed by scattering neutrons off the nuclei; however, systematically clean, high-statistics data can be challenging to obtain, resulting in a relatively sparse collection of data worldwide. High-quality, repeatable and verifiable measurements are a priority to improve the confidence in the interpretation of the CEvNS searches. The COHERENT collaboration addresses this with a dedicated neutron scattering facility with a pulsed, collimated, quasi-monoenergetic, tunable beam at the tandem Van de Graaff accelerator at TUNL~\cite{Akimov:2017ade,Rich:2017lzd,Akimov:2021ggw}.

\section{SUMMARY}

At the time of this writing, COHERENT in Neutrino Alley at the SNS has made the first and second detections of CEvNS, on one heavy and one lighter nucleus. This first measurement has enabled new constraints of BSM physics, pushed sub-GeV dark matter limits to beyond the cosmological expectation, and opened a window for new nuclear physics measurements.  Plans are underway to fill in more target isotopes, first with Ge and Na,  to evaluate the $N^2$ dependence of the cross section and to make yet more stringent tests of the SM.   Neutrino absolute flux uncertainties are dominant and common to all detectors.   A heavy water detector will improve these flux uncertainties via CC $\nu_e$ measurement on deuterons, for which the cross section is well understood. Additional QF measurements will further reduce systematic uncertainties.  CEvNS is not the only accessible neutrino interaction at the SNS: inelastic CC and NC cross sections,  with particular relevance for supernova neutrinos, are also planned. COHERENT detectors are sensitive and competitive in the search for accelerator-produced DM in the sub-GeV mass range, as well as possibly other BSM particles.   A  neutrino flux upgrade at the SNS is underway, which will both deepen and broaden the program. 

\section*{DISCLOSURE STATEMENT}
 The authors are not aware of any affiliations, memberships, funding, or financial holdings that
might be perceived as affecting the objectivity of this review. 

\section*{ACKNOWLEDGMENTS}
The authors wish to thank the COHERENT collaboration and Oak Ridge National Laboratory for support.  COHERENT is is funded by multiple sources, including the U.S. Department of Energy, the National Science Foundation and the Russian Foundation for Basic Research.

\bibliographystyle{ar-style5.bst}
\bibliography{refs}

\end{document}